\documentclass[preprint2]{aastex}
\shorttitle{Modeling the Time Variability of S82 Quasars}
\shortauthors{MacLeod et al.}
\begin{document}

\title{Modeling the Time Variability of SDSS Stripe 82 Quasars as a
  Damped Random Walk}

\author{C.~L.~MacLeod\altaffilmark{1,2},
\v{Z}.~Ivezi\'{c}\altaffilmark{2},
C.~S.~Kochanek\altaffilmark{3,4}, 
S.~Koz{\l}owski\altaffilmark{3},
B.~Kelly\altaffilmark{5,6}, 
E.~Bullock\altaffilmark{2}, 
A.~Kimball\altaffilmark{2}, 
B.~Sesar\altaffilmark{2}, 
D.~Westman\altaffilmark{2,7}, 
K.~Brooks\altaffilmark{2},
R.~Gibson\altaffilmark{2}, 
A.~C.~Becker\altaffilmark{2}, and 
W.~H.~de~Vries\altaffilmark{8}}

\altaffiltext{1}{cmacleod@astro.washington.edu}
\altaffiltext{2}{Department of Astronomy, University of
  Washington, Box 351580, Seattle, WA 98195} 
\altaffiltext{3}{Department of Astronomy, The Ohio State University, 140 West
  18th Avenue, Columbus, OH 43210} 
\altaffiltext{4}{The Center for Cosmology and Astroparticle Physics, The Ohio
  State University, 191 West Woodruff Avenue, Columbus, OH 43210} 
\altaffiltext{5}{Hubble Fellow}
\altaffiltext{6}{Harvard-Smithsonian Center for Astrophysics, 60 Garden St, 
Cambridge, MA 02138}
\altaffiltext{7}{James Cook Univ., Centre for Astronomy, Townsville, QLD 4811,
Australia}
\altaffiltext{8}{University of California, One Shields Ave,
  Davis, CA 95616}

\begin{abstract}
  We model the time variability of $\sim$9,000
  spectroscopically confirmed quasars in SDSS Stripe 82 as a damped
  random walk. Using 2.7 million photometric measurements collected over 10 years, we confirm
  the results of Kelly et al.~(2009) and Koz{\l}owski et al.~(2010) that this
  model can explain quasar light curves at an impressive fidelity
  level (0.01-0.02 mag). The damped random walk model provides a
  simple, fast [O($N$) for $N$ data points], and powerful statistical
  description of quasar light curves by a characteristic time scale ($\tau$)
  and an asymptotic rms variability on long time scales
  ($SF_{\infty}$).  We searched for correlations
  between these two variability parameters and physical parameters
  such as luminosity and black hole mass, and rest-frame wavelength. 
  Our analysis shows $SF_{\infty}$ to increase with
  decreasing luminosity and rest-frame wavelength as
  observed previously, and without a correlation with redshift. We find a
  correlation between $SF_{\infty}$ and black hole mass with a power
  law index of 0.18$\pm$0.03, independent of the anti-correlation with
  luminosity. We find that $\tau$ increases with increasing wavelength
  with a power law index of 0.17, remains nearly constant with redshift
  and luminosity, and increases with increasing black hole mass with
  power law index of 0.21$\pm$0.07.  The amplitude of variability is 
  anti-correlated with the Eddington ratio, which suggests a scenario where
  optical fluctuations are tied to variations in the accretion
  rate. However, we find an additional dependence on luminosity and/or
  black hole mass that cannot be explained by the trend with Eddington
  ratio. The radio-loudest quasars have systematically larger
  variability amplitudes by about 30\%, when corrected for the other
  observed trends, while the distribution of their characteristic time scale is  
  indistinguishable from that of the full sample. We do not detect any
  statistically robust differences in the characteristic time scale and variability
  amplitude between the full sample and the small subsample of quasars 
  detected by ROSAT.  Our results provide a simple quantitative framework for
  generating mock quasar light curves, such as currently used in LSST
  image simulations.
\end{abstract}

\maketitle

\section{Introduction}
The optical variability of quasars has been recognized since they were
 first identified (Matthews \& Sandage 1963). Indeed, most quasars are variable
 ($\sim$90\% at the 0.03 mag rms level; Sesar et al.~2007), and the variations
 in brightness are aperiodic and on the order of 20\% on
time scales of months to years (e.g., Hook et al.~1994; Vanden Berk et
al.~2004). Furthermore, the smooth power spectra suggest a chaotic, or 
stochastic, origin for the variability. A range of models have been 
advanced to describe quasar variability, including supernova 
bursts, microlensing, and accretion disk instabilities (Aretxaga et
al.~1997; Hawkins~1993; Kawaguchi et al.~1998; Tr{\`e}vese \&
 Vagnetti~2002). These models are discussed and compared in Hawkins~(2007). 
Reverberation mapping studies (e.g., Peterson et al.~2005) show that the 
broad emission lines respond to continuum fluctuations, therefore providing 
strong evidence that the variability is intrinsic to the
quasars. A number of studies have utilized standard accretion disk models to 
demonstrate that the optical-UV variability of quasars could be driven by a variable
accretion rate (e.g., Pereyra et al.~2006; Li \& Cao 2008; Liu et al.~2008).
Blackburne \& Kochanek (2010) find evidence in the light curves of microlensed quasars 
that the optical variability is caused by a change in the effective area of the accretion disk. 

Recently, Kelly et al.~(2009, hereafter KBS09) proposed a model where the optical
variability is described by a damped random walk
(a self-correcting term added to a random walk model that acts to push
any deviations back towards the mean value). They proposed that the
variability time scale might be identified with the thermal time scale
of accretion disks, as also proposed by Collier \& Peterson~(2001). A thermal
origin of the variability would explain why quasars become bluer as they brighten (e.g.,
Giveon et al.~1999; Tr{\`e}vese et al.~2001; Geha et al.~2003). 

Although the physical causes have yet to be proven, 
it has been established by KBS09 and
Koz{\l}owski et al.~(2010a, hereafter Koz10) that a damped random
walk can statistically explain the observed light curves of quasars. 
Using 100 well-sampled single-band light curves compiled from the literature, 
KBS09 show that 
this stochastic process is capable of modeling complex quasar light curves at an impressive
fidelity level (0.01-0.02 mag).  
Koz10 applied the model to the OGLE light curves (Udalski et al.~1997; Udalski et
al.~2008) of mid-infrared-selected quasars
behind the Magellanic Clouds from Koz{\l}owski \& Kochanek (2009). 
Their analysis shows that this stochastic model is robust
enough to efficiently select quasars from other variable sources (see Schmidt 
et al.~2010 for a different method of selecting quasars based on variability).
The model has only three free parameters:
the mean value of the light curve, the driving amplitude of the stochastic process, and the 
damping time scale. The predictions are only statistical, and the
random nature reflects our uncertainty about the details of the physical processes.

Instead of applying a model to observed light curves for
individual quasars, numerous studies have looked at the ensemble 
variability of quasars, particularly in samples where individual light
curves are not available. Significant progress in the description of quasar
variability has been made by employing the Sloan Digital Sky Survey (SDSS) 
data (Vanden Berk et al.~2004, hereafter VB04; Ivezi\'{c} et
al.~2004, hereafter I04; de Vries et al.~2005; Wilhite et al.~2005,
2006, 2008; Sesar et al.~2006).   
For example, the size and quality of the sample analyzed
by VB04 (two-epoch photometry for 25,000
spectroscopically confirmed quasars) allowed them to constrain how
quasar variability in the rest frame optical/UV regime depends upon
rest-frame time lag (up to $\sim$2 years), luminosity, rest
wavelength, redshift, the presence of radio and X-ray emission, and
the presence of broad absorption line systems. 
Using repeated SDSS photometric observations, Wilhite
et al.~(2008) confirmed the result of Wold et al.~(2007) that
variability is correlated with black hole mass, and show that this is independent of the
anti-correlation between variability and luminosity established by many
studies. This led them to 
suggest that the amplitude of variability may be driven by the quasar's 
Eddington ratio, implying differences in accretion rate. 

These studies typically quantify the observed optical variability of quasars using a
structure function (SF) analysis (see also Hughes et al.~1992; Collier
\& Peterson 2001; Bauer et al.~2009; Koz{\l}owski et al.~2010b), where
the SF is the root-mean-square (rms) magnitude difference as a
function of the time lag ($\Delta t$) between measurements. This
autocorrelation-like function is less sensitive to aliasing and other
time-sampling problems than a power spectral distribution. 
By studying the magnitude difference distribution for appropriately
chosen subsamples with fixed values of absolute i-band magnitude
($M_i$), rest-frame time lag ($\Delta t_{RF}$, in days), and
wavelength ($\lambda_{RF}$, in \AA), the mean dependence of the SF on
these quantities was inferred by I04 to be  
\begin{equation}
SF_{model}= A[1 + B\,M_i] \left(\frac{\Delta
  t_{RF}}{\lambda_{RF}}\right)^C \rm mag,
\label{eq:I04}
\end{equation} 
with $A=1.00 \pm 0.03$, $B=0.024 \pm 0.04$, and $C=0.30 \pm 0.05$. 
A qualitatively similar result was obtained by VB04. 
Koz{\l}owski et al.~(2010b), in the first large study of the mid-IR 
structure functions of quasars, also found lower variability for
higher luminosities and longer wavelengths, but the temporal slope
of the ensemble structure functions were significantly steeper than in 
the optical.
In addition, there is evidence for a turnover in the SF on long time lags 
(I04; Rengstorf et al.~2006; Wold et al.~2007). 
Studies by de Vries et al.\ (2005) and Sesar et al.\ (2006) using SDSS
combined with earlier Palomar Observatory Sky Survey measurements for
40,000 SDSS quasars constrained quasar continuum
variability on time scales of 10 to 50 years in the observer's frame.  
They report that the characteristic time scale, which in this context is the 
time lag above which the SF flattens to a constant value,
is of order 1 year in the quasar rest frame.
Using a shot-noise light curve model, de Vries et al.~(2005) found evidence
for multiple variability time scales in long-term ensemble variability
measurements, while Collier \& Peterson (2001) found a wide range of
different time scales in their analysis of individual light curves,
and even evidence for multiple time scales in a single active galactic
nucleus (AGN). 

These analyses of ensemble variability are based on a fundamental
assumption that photometric observations at two epochs for a large
number of quasars will reveal the same statistical properties as
well-sampled light curves for individual objects.  This assumption has
been tested by MacLeod et al.~(2008) using light 
curves for spectroscopically confirmed quasars observed roughly 50 times
over 8 years in SDSS Stripe 82 (S82). They found that
while the mean SF for individual sources is consistent with
Eq.~\ref{eq:I04}, the contribution of the mean trends to the observed
dispersion in variability properties is minor compared to an intrinsic
stochasticity of unknown origin. Further investigation of this
stochastic behavior is one of the main goals of this study. 

In order to better understand the relationship between the two types
of data analyses (individual versus ensemble quasar variability), and to begin linking to
physical models, we apply the damped random walk model 
to the $ugriz$ light curves of $\sim$9,000 spectroscopically confirmed SDSS
S82 quasars.  This large sample greatly benefits from the robust, accurate, 
five-band SDSS photometry. We estimate the variability parameters following
Koz10, who demonstrated that their approach is more statistically powerful than
the forecasting methods used by KBS09. We also note that both the
Koz10 and KBS09 approaches are much faster than that used by Schmidt et
al.~(2010), requiring only O($N$) rather than O($N^2$) operations to
determine the model parameters for a light curve with $N$ data
points. 

In Section~\ref{sec:methodology}, we describe the model, define our variability
parameters, and demonstrate their relationship to those utilized in previous studies. 
In Section~\ref{sec:data}, we introduce the S82 data set and outline our initial light curve selection. 
In Section~\ref{sec:results}, we present the best-fit variability
parameters for our final sample of light curves and 
estimate their scatter due to the limited time sampling of SDSS.
We also estimate the sensitivity of our results to variations in the 
slope of the model power spectrum on long time scales. 
In Section~\ref{sec:trends}, we describe the relationship between the long-term variability parameters and 
physical parameters such as wavelength, absolute magnitude, black
hole mass, and Eddington ratio. 
Using these results, we also provide a prescription for simulating
mock quasar light curves. 
In Section~\ref{sec:radio}, we explore the variability properties of
subsamples detected at radio and X-ray wavelengths.
Finally, we summarize our results in Section~\ref{sec:discussion}.

\section{Methodology}
\label{sec:methodology}
We model the time variability of quasars as a stochastic process
described by the exponential covariance matrix
\begin{equation}
        S_{ij} = \sigma^2 \exp(-|t_i-t_j|/\tau)
 \label{eqn:cfunc}
\end{equation}
between times $t_i$ and $t_j$.  As detailed by KBS09 and Koz10,
this corresponds to a damped random walk (more specifically, an
Ornstein-Uhlenbeck process) with a damping time scale $\tau$, also
called the characteristic time scale, and a long-term standard
deviation of variability\footnote{The $\sigma$ used here 
  is related to the $\sigma$ used in KBS09 ($\sigma_{KBS}$) and the
  parameter $\hat{\sigma}$ used in Koz10 as 
  $\sigma_{KBS} = \hat{\sigma} = \sigma\sqrt{2/\tau}$.} 
$\sigma$. Following Koz10, we model the light curves and estimate
  the parameters and their uncertainties using the method of 
Press et al.~(1992), its generalization 
in Rybicki \& Press (1992), and the fast computational
implementation described in Rybicki \& Press (1994).  Koz10 
demonstrate that this approach is more statistically powerful than
the forecasting methods used by KBS09, while still
having computation times scaling linearly with the number of data
points. 

\subsection{Structure Function for the Damped Random Walk Model}
\label{sec:SF}
For our analysis, we express the long-term variability in terms of the
structure function (SF) in order to relate to our previous studies and
to those of two-epoch samples. 
The first order SF is $SF^2(\Delta t) =
2\sigma^2[1-ACF(\Delta t)]$ (e.g., Hughes et
al.~1992), where the autocorrelation function for a damped random walk is $ ACF(\Delta t) = \exp{(-|\Delta t|/\tau)}$ 
(Eq.~A6 in KBS09). This ACF results in the structure function\footnote{The functional form of SF$(\Delta
  t)$ fit to the long-term SDSS-POSS data in Sesar et al.~(2006), see their Eq.~5, is similar but
not identical to the functional form given by Eq.~\ref{eq:sfdt}.}: 
\begin{equation}
  SF(\Delta t) = SF_{\infty}(1-e^{-|\Delta t|/\tau})^{1/2}.
\label{eq:sfdt}
\end{equation}
Asymptotic values of the SF at large and small $\Delta t$ are 
\begin{eqnarray} 
  SF(\Delta t >> \tau) \equiv SF_{\infty} = \sqrt{2}\sigma   \nonumber \\
  SF(\Delta t << \tau) = \sigma \sqrt{\frac{2|\Delta t|}{\tau}} = SF_{\infty}\sqrt{\frac{|\Delta t|}{\tau}}
\label{eq:sfinf}
\end{eqnarray}
The form $SF\propto |\Delta t|^{\beta}$ is equivalent to a power
spectral distribution $PSD \propto f^{\alpha}$, where
$\alpha=-2\beta-1$ (see Appendix of Bauer et al.~2009; KBS09).
The SF at small time lags is therefore equivalent to a power
spectral distribution $PSD \propto f^{-2}$.
We adopt $SF_{\infty}$ and $\tau$ as our two main variability
model parameters.

\subsection{Model Light Curves}
\label{sec:modelLCs}
Equipped with a statistical description of quasar variability, 
we generate well-sampled light curves 
in order to 1) demonstrate the relationship between our variability parameters 
and the traditional SF analyses of many previous works, and 2) 
to estimate the systematic effects that the sampling rate and light 
curve length have on the fitted parameters. The latter is especially 
important because the S82 data are fairly sparse. 
As shown below and in Section~\ref{sec:tests}, these indeed have 
a large impact. 

\begin{figure*}[p]\centerline{
\includegraphics[width=5.2in]{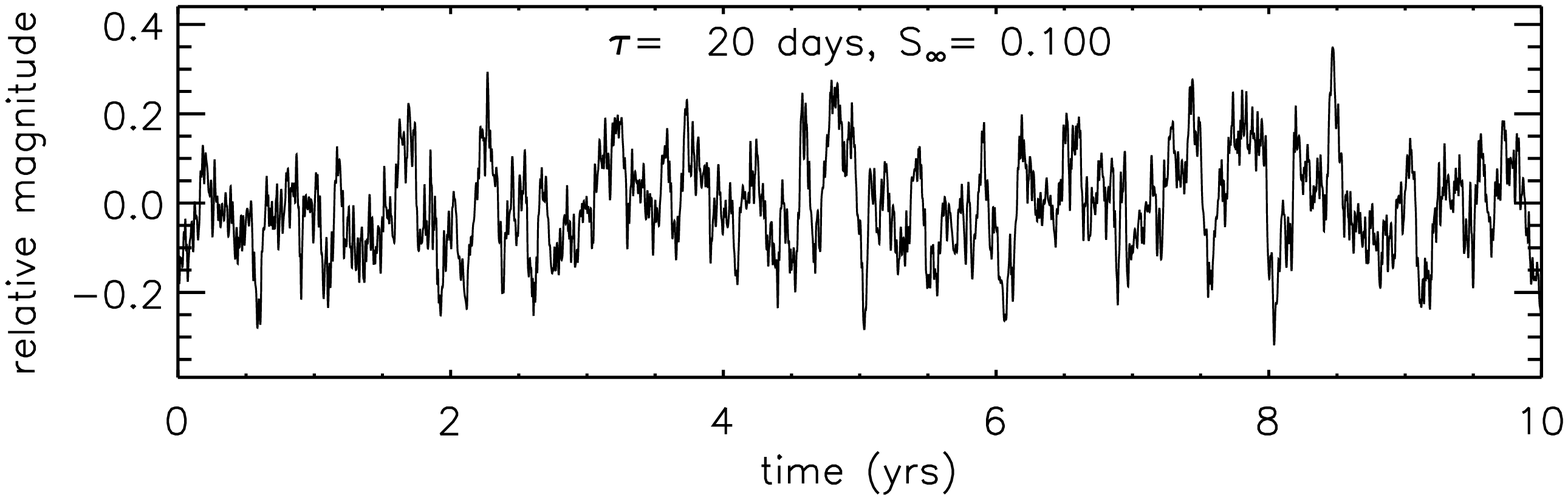}}\centerline{
\includegraphics[width=5in]{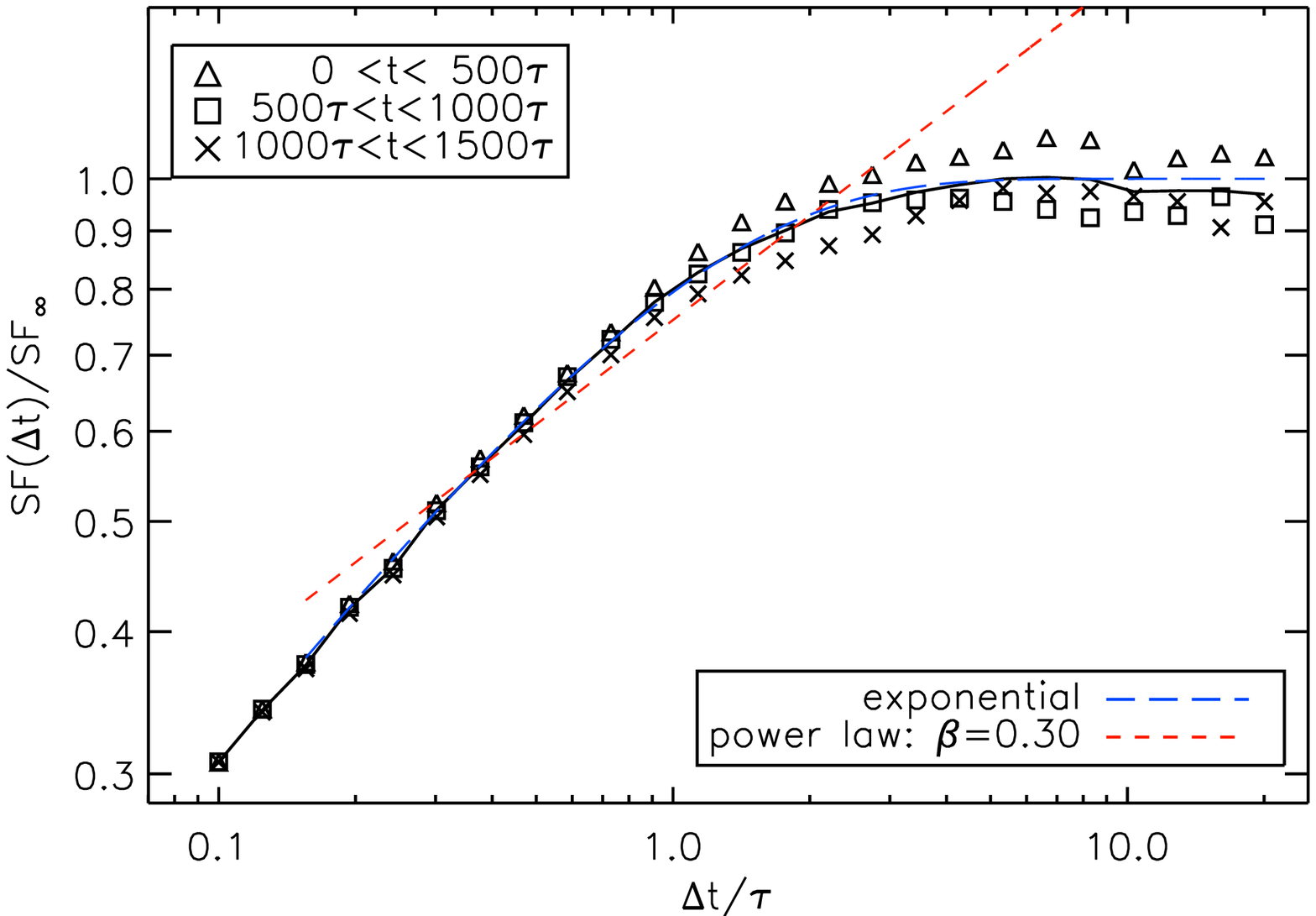}}
\caption{The top panel shows a segment of a simulated light curve with
  $\tau=20$ days and  $SF_{\infty}=0.10$ mag. In the bottom panel, 
  triangles, squares, and crosses represent the SF computed for the
  first, second, and third sections of the total light curve,
  respectively (each has a length of $500\tau$). 
  The increased scatter in $SF(\Delta t)$ at large time lags  
  is due to the finite length of each light curve section -- 
  the scatter decreases when using the entire light curve length of $1500\tau$
  (as shown by the solid line). 
  The short-dashed line is a power law fit, $SF(\Delta t)\propto \Delta
  t^\beta$ with $\beta=0.3$, to the data points with $0.15<\Delta
  t/\tau<3$. The long-dashed line is the true function 
  $SF = SF_{\infty}[1-e^{-\Delta t/\tau}]^{1/2}$.}
\label{fig:splittest}
\end{figure*}

A light curve is generated using only three input parameters: 
$SF_{\infty}$, $\tau$, and the mean value of the light curve, $\mu$.
The magnitude $X(t)$ at a given timestep $\Delta t$ from a 
previous value $X(t-\Delta t)$ is drawn from a normal
distribution with a mean and variance given by 
\small
\begin{eqnarray}
E(X(t)|X(t-\Delta t)) = e^{-\Delta t/\tau}X(t-\Delta t)+ \nonumber \\ \mu(1-e^{-\Delta t/\tau})
\nonumber \\
Var(X(t)|X(t-\Delta t)) = 0.5(SF_{\infty})^2(1-e^{-2\Delta t/\tau})
\label{eq:car1sim}
\end{eqnarray}
\normalsize
(Eqs.~A4 and A5 in KBS09). The asymptotic variance of the time series
is then $0.5(SF_{\infty})^2$.   
The top panel of Figure~\ref{fig:splittest} shows a segment of a well-sampled light
curve generated using $\tau=20$ days, $SF_{\infty}=0.14$ mag, and a
time sampling of $0.1(\Delta t/\tau)$. 
The structure function, $SF(\Delta t)$, is computed by collecting 
the differences in magnitude for all points in the
light curve separated by a given time lag, $\Delta t$. The distribution of magnitude
differences ($\Delta m$) is Gaussian by construction, and 
the rms of the $\Delta m$ distribution is the $SF$ value for that time lag. 

When fitting values for $SF_{\infty}$ and $\tau$ for a given light
curve, the length of time that it spans plays an important role. 
For example, in Figure~\ref{fig:splittest}, the SF computed for the 
full light curve length of $1500\tau$ (82 years) is much smoother than that
computed for three equal sections, each spanning 27 years. 
Therefore, the determination of variability parameters for S82 quasars
will be affected by their light curve lengths, which are typically $10\tau$.
When the light curve is too short, it is easy to overestimate $\tau$ because 
there is no information on the time scale of the break (where 
$SF(\Delta t)$ flattens to $SF_{\infty}$).  The model will reproduce 
the observed variance in the light curve by overestimating 
$SF_{\infty}\simeq (SF_{\infty})_{true}(\tau/\Delta t)^{1/2}$, and therefore 
$\hat{\sigma}=SF_{\infty}/\sqrt{\tau}$ is the more robustly 
estimated model parameter when $\tau$ cannot be well-determined.

\subsection{Comparison with Published Work}
\label{sec:comparison}
Before interpreting the form of $SF(\Delta t)$, we summarize the major 
differences between our S82 analysis and previous studies based  
on ensemble structure functions.
The ensemble structure function is computed using only a few observations 
of many quasars, combining all magnitude differences to find $SF(\Delta t)$. 
Using an ensemble approach is beneficial because it enables one to
constrain the average variability properties 
when it is difficult to constrain such quantities for individual
quasars. Indeed, even with well-sampled light curves, spurious breaks
in the individual SFs are common (Emmanoulopoulos, McHardy, \& Uttley 2010).
However, in previous works (e.g., I04; de Vries et al.~2005), 
the characteristic time scale is defined as the time lag at which the ensemble $SF(\Delta t)$ 
flattens to a constant value, and thus it represents some complex
average over the intrinsic $\tau$ distribution. 
In contrast, by applying a stochastic model to the individual S82 light 
curves, we are relatively insensitive to time sampling issues (for
details, see KBS09), and we obtain a model fit for every quasar in
each filter described by the parameters $SF_{\infty}$ and
$\tau$. 

A power law fit to $SF(\Delta t)$ has previously been a common way
to describe $SF(\Delta t)$, and even to reject certain classes of models (e.g., Kawaguchi et al.~1998; VB04; I04). 
However, Figure~\ref{fig:splittest} shows that the best-fit power law index 
is extremely sensitive to the fitted range of $\Delta t/\tau$. 
For example, for $0.1<\Delta t/\tau<1$, 
$SF(\Delta t)$ is well fit by a power law with an index of $\sim$0.5, while 
for $0.15<\Delta t/\tau<3$, we obtain a strongly biased power law index of $0.3$. 
Furthermore, each quasar has its own values of $\tau$ and $SF_{\infty}$,
and the ensemble structure function is a convolution
 of the individual structure functions with their distribution in parameters, 
\small
  \begin{equation}
      SF(\Delta t) = \int {\rm d}\tau {\rm d}SF_{\infty}{ {\rm d}^2 n
	\over {\rm d}\tau {\rm d}SF_{\infty} }
      SF(\Delta t| \tau, SF_{\infty}),
  \label{eq:ensSF}
  \end{equation}
\normalsize
where $SF(\Delta t| \tau, SF_{\infty})$ (Eq.~\ref{eq:sfdt}) is the structure function at
time $\Delta t$ for a quasar with variability parameters $\tau$ and
$SF_{\infty}$. Hence, the ensemble structure function is only indirectly related to the
structure function for any particular quasar,  and results based on fitting a power 
law to observed ensemble structure functions should be interpreted
with caution.

\section{The SDSS Stripe 82 Quasar Data Set}
\label{sec:data}
The Sloan Digital Sky Survey (SDSS, York et al.~2000) provides
homogeneous and deep ($r < 22.5$) photometry in five passbands
($ugriz$, Fukugita et al.~1996; Gunn et al.~1998; Smith et al.~2002)
accurate to 0.02 mag, of almost 12,000 deg$^2$ in the Northern
galactic cap, and a smaller, but deeper, survey of 290~deg$^2$ in the
Southern galactic hemisphere. For this 290~deg$^2$ area known as Stripe
82 (S82), there are on average more than 60 available
epochs of observations. These data were obtained in yearly ``seasons'' about
2-3 months long over the last decade, and the cadence effectively
samples time scales from days to years. 
The light curve lengths are effectively shorter than the actual 
period of the survey because the better-sampled supernova observations 
begin about 5 years into the survey.  
Because some observations were obtained in non-photometric
conditions, improved calibration techniques have been applied to SDSS
S82 data by Ivezi\'{c} et al.~(2007) and Sesar et al.~(2007), and we
use their  results. 
For these data, photometric zero-point errors are 0.01--0.02 mag. 

We have compiled a sample of 9,275 spectroscopically confirmed quasars in S82 
with re-calibrated $ugriz$ light curves (see also Bhatti et al.~2010).
Most (8,974) of these are in the SDSS Data Release 5 (DR5) Quasar
Catalog (Schneider et al.~2007), and the remaining are newly confirmed
DR7 (Abazajian et al.~2009)  quasars.
Summed over all bands and epochs, the data set includes 2.7 million
photometric measurements. For 41\% of the sample, the random photometric
errors are smaller than 0.03 mag. Only 1\% have errors $\geq 0.1$ mag in 
$g$, $r$, and $i$, and 2.4\% have errors exceeding 0.25 mag in $u$ and $z$ filters. 
In MacLeod et al.~(2010, in prep.), these light curves, as
well as a much larger sample of quasars with two SDSS epochs selected
from 12,000 deg$^2$ of the sky, will be made publicly
available. We adopt the K-corrected i-band absolute magnitudes from
Schneider et al.~(2007), and virial black hole masses and
bolometric luminosities where available from Shen et
al.~(2008). The Shen et al. masses were estimated from emission line
widths (H$\beta$ for $z<0.7$, MgII for $0.7 < z < 1.9$, and CIV for $z
> 1.9$). However, we note that at low spectroscopic signal-to-noise,
black hole masses tend to be overestimated (Denney et al.~2009). 

\subsection{Initial Light Curve Selection}
\label{sec:cuts}
\begin{figure*}[p]
\centering
\includegraphics[width=5in]{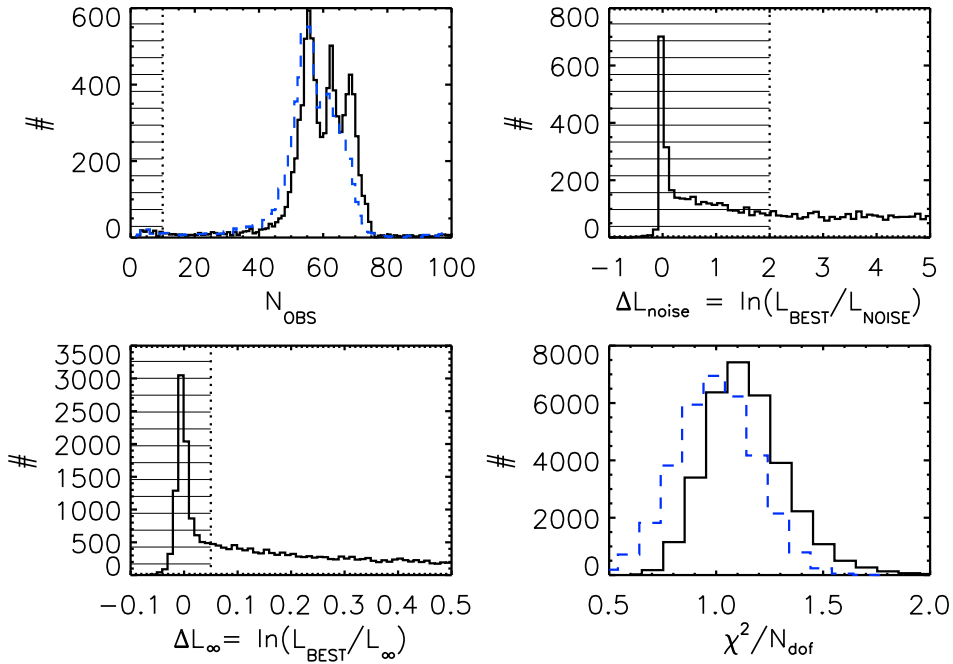}
\caption{Initial light curve selection. \emph{Top-left}: the distribution of the number
  of observations per light curve for the $r$ (solid) and
  $u$ (dashed) bands. \emph{Top-right}: distribution of $\Delta
  L_{noise}$. We define light curves with $\Delta
  L_{noise}\leq 2$ to be more consistent with uncorrelated noise
  rather than our model. \emph{Bottom-left}: distribution of $\Delta
  L_{\infty}$; light curves with $\Delta
  L_{\infty}\leq 0.05$ likely have run-away time scales. In this panel
  and the previous panel, the x-axes are truncated at $0.5$ and $5$
  respectively, but the histograms continue to greater values. 
  \emph{Bottom-right}: distribution of 
  $\chi^2$ per degrees of freedom ($N_{dof}$) for the damped random
  walk model (solid line). The expected
  Gaussian distribution based on $N_{dof}$ is also shown (dashed).   
  The hashed region in each
  panel shows the values rejected from our final sample. }
\label{fig:cuts}
\end{figure*}

The damped random walk model was fit to all available $ugriz$ light
curves for $9,275$ S82 quasars. Summed over 5 bands, there are
$46,375$ best-fit values of the characteristic (damping) time scale
$\tau$ and long-term  structure
function $SF_{\infty}$. 
For further analysis, we select light curves that satisfy the following criteria:
\begin{enumerate}
\item First, we remove light curves with fewer than 10 observations.
  The top-left panel of Figure~\ref{fig:cuts} shows the distribution of 
  the number of observations ($N_{obs}$) per light curve before this
  cut, which reduces our sample to $45,814$. 
  At a given $N_{obs}$, the distribution of the
  ratio of light curve length to $\tau$ is similar to that at all other
  values of $N_{obs}$. Therefore, any systematic effects of $N_{obs}$ on 
  derived parameter distributions should be small. 
\item We then require that the stochastic model must provide a better fit than 
  uncorrelated noise. Following Koz10, we select light curves with a
  likelihood improvement over simply broadening the measurement errors of 
  $\Delta L_{noise} \equiv \ln{(L_{best}/L_{noise})}>2$, where $L_{best}$ is the
  likelihood of the stochastic model and $L_{noise}$ is that for the
  noise solution where $\tau \rightarrow 0$. 
  The top-right panel of Figure~\ref{fig:cuts} shows the distribution
  of $\Delta L_{noise}$ before this cut, which removes 14\% of $u$ and
  $z$ light curves, whose photometric errors are larger.
  About 7\% of our light curves (82\% of which are $u$ or $z$ band) are removed in this 
  step\footnote{A detailed analysis of the use of this variability
  model to select candidate quasars will be presented elsewhere (Brooks et al., in prep.).}, 
  reducing our total to $42,623$.  
\item Finally, we remove cases where $\tau$ is 
  merely a lower limit due to the length of the light curve. 
  We define 
  $\Delta L_{\infty} \equiv \ln{(L_{best}/L_{\infty})}$, 
  where $L_{\infty}$ is the likelihood that $\tau \rightarrow \infty$, 
  indicating that the light curve length is too short to accurately
  measure $\tau$. 
  The bottom-left panel shows a peak at $\Delta L_{\infty}=0$, and we
  exclude these objects by requiring that  $\Delta L_{\infty}>0.05$. 
  Most (95\%) of the rejected light curves have lengths 
  $<\tau$; 76\% have $\tau \geq 10^4$ days, and
  64\% have $SF_{\infty} \geq 1$ mag. The latter is due to the fact
  that as the $\tau$ value becomes long and uncertain, the model will
  necessarily overestimate $SF_{\infty}$ in order to keep the overall
  light curve rms fixed (see Section~\ref{sec:modelLCs}). 

  The rejected light curves tend to be higher redshift quasars because
  stronger time dilation leads to shorter rest-frame light curve
  lengths, making it increasingly difficult to constrain long
  rest-frame $\tau$. This criterion removes 22\% of our sample, leaving 
  a total of $33,112$ values of $\tau$ and $SF_{\infty}$. Because we
  are limited by the duration of the S82 survey, this is a significant
  loss, and therefore our final $\tau$ distribution is biased
  low, but the bias should not be significant considering our results
  in Section~\ref{sec:tests}. 
\end{enumerate}

The resulting light curves are well-fit by the stochastic model, as
can be seen from the distribution of $\chi^2/N_{dof}$ 
shown in the bottom-right panel in Figure~\ref{fig:cuts}, 
where $N_{dof}$ is the number of degrees of freedom.  
The expected Gaussian distribution with rms 
$=\sqrt{2/N_{dof}}$ is also shown in the panel, where we have 
averaged over the $N_{dof}$ distribution of the light curves.
The observed distribution is centered at $\chi^2/N_{dof}=1.1$. This
difference is some combination of errors in the estimated errors,
outliers in the light curves, and any poorly modeled physics. 
Koz10 noted a similar difference in their analysis of OGLE light
curves. Only 5\% of the light curves have $\chi^2/N_{dof}>1.5$, 
confirming that most quasars are variable (at the $\Delta L_{noise}>2$
level), and that a damped random walk is a good description of quasar variability.  

\section{Variability Parameters for Stripe 82 Quasars}
\label{sec:results}

\subsection{Observed Distributions}
\label{sec:bestfits}
We assume the variability to be intrinsic to the quasars and convert
the time scales to the rest frame (dividing by $(1+z)$) before further analysis.
The long-term structure function should be independent of redshift 
other than through evolution in physical parameters, and this view is
confirmed by KBS09. 
Figure~\ref{fig:pardist} shows the distributions in $SF_{\infty}$ and
rest-frame $\tau$ found for the S82 quasars. 
If we consider only the brighter ($i < 19$) quasars, based on the
best-fit mean magnitude, the distributions are generally similar but
biased towards lower asymptotic amplitudes (peaked at 0.12 mag). 
The distributions are consistent with what was found in KBS09 and
Koz10; however, in these studies, there are not as many objects with
runaway ($\tau \rightarrow \infty$) time scales because of the
improved time sampling of the OGLE survey and many of the light curves
in the KBS09 sample. Also, the observed-frame $\tau$ distribution
lacks many of the short time scales observed in Koz10; this is
likely due to either the better time sampling of the OGLE light
curves or unrecognized stellar contamination in their sample.

The bottom panel of Figure~\ref{fig:pardist} shows that the best-fit
variability parameters are highly correlated with each other,
indicating that quasars with larger asymptotic amplitudes of
variability also have longer characteristic time scales. We fit a
power law slope of $1.3 \pm 0.01$ dex/dex for all 33,112 data points,
and this trend persists within each $ugriz$ band as well. Note that 
a correlation between $\tau$ and $SF_{\infty}$ is expected even if
$\tau$ is independent of the driving amplitude of short-term 
variations, $\hat{\sigma}=SF_{\infty}/\sqrt{\tau}$. Since the power
law slope of $1.3$ is steeper than that expected if $\tau$ and
$\hat{\sigma}$ are independent ($\sim$0.2, see below), the time scale
must also be correlated with the amplitude of short-term variability.
We confirm that this correlation is intrinsic, rather than an artifact of
short light curve lengths, in Section~\ref{sec:tests}. 

The approximate values for $\tau$ and $SF_{\infty}$ can often be
guessed from light curves by simple visual inspection. 
Figure~\ref{fig:LCeg1} shows representative light curves from several 
regions of $\tau-SF_{\infty}$ space, indicated by stars in
Figure~\ref{fig:pardist}. The weighted average of all model light
curves consistent with the data is also shown, and the ``error 
snake'' is the $\pm 1\sigma$ range of those light curves about this mean.
The top panel shows a light curve with a relatively short $\tau$ -- this
is due to the large amount of variability within each season (i.e.\ each
grouping of points), as compared to that in the second panel, which
shows a curve with a much larger $\tau$.  The third panel shows a
light curve with a relatively low $SF_{\infty}$, while the bottom
panel shows one with a larger $SF_{\infty}$, as can be seen by the
larger difference in median $r$ band magnitude between about 10 and 11
years. 
Note the outlier in the third panel: outliers such as these are
generally responsible for higher values of $\chi^2/N_{dof}$. 
The (data$-$model) brightness difference  provides a convenient way to identify 
outliers, and they may be an indicator of variability behavior not captured by the 
damped random walk model. However, they may also be caused by 
occasional non-Gaussian photometric errors and thus their 
analysis requires a detailed and careful study (e.g., by utilizing
control samples of appropriately chosen nearby non-variable stars). 
Since the model provides satisfactory fits for the overwhelming
majority of objects (see Section~\ref{sec:cuts}), we do not further
investigate such outliers in this work. 
We have also searched for periodic signals in observed light curves 
and did not find any convincing cases. This analysis is summarized
in the Appendix.

\subsection{The Effect of S82 Time Sampling and Estimate of Fitting Errors}
\label{sec:tests}
There are three contributions to the scatter in the best-fit
variability parameters $\tau$ and $SF_{\infty}$. First, the fitting
errors, including those due to insufficient time sampling and light
curve length, will introduce some scatter. Second, trends with
physical parameters (see Section~\ref{sec:trends}) result in a certain
distribution width. There may also be scatter due to other sources of
variability that are not captured by the model, such as flares, or
other activity related to radio emission, for example. Here, we carry
out two tests in order to understand how the best-fit parameters are
affected by the limited data sampling of S82.  

\begin{figure*}[p]
\centerline{\includegraphics[width=4in]{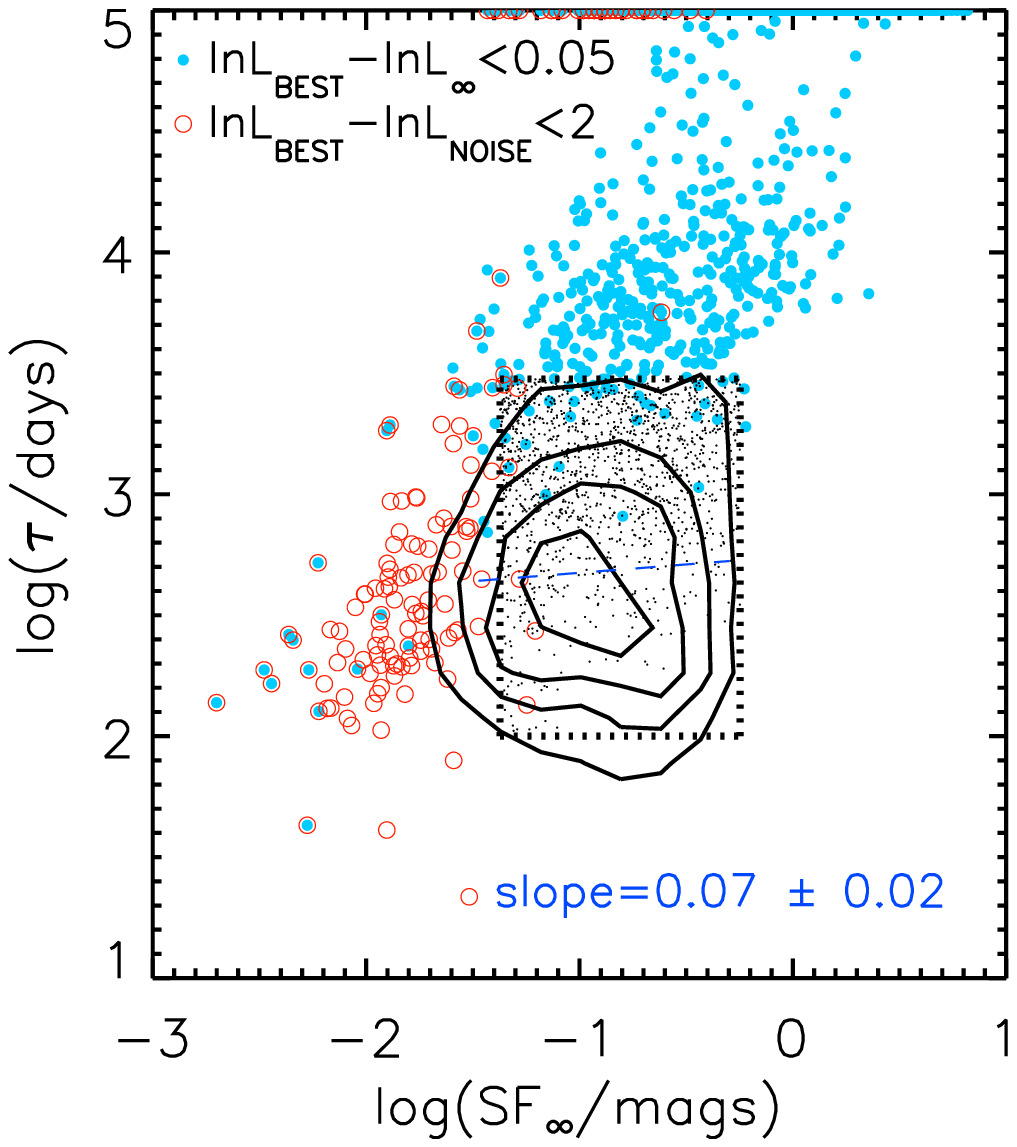}
  \includegraphics[width=4in]{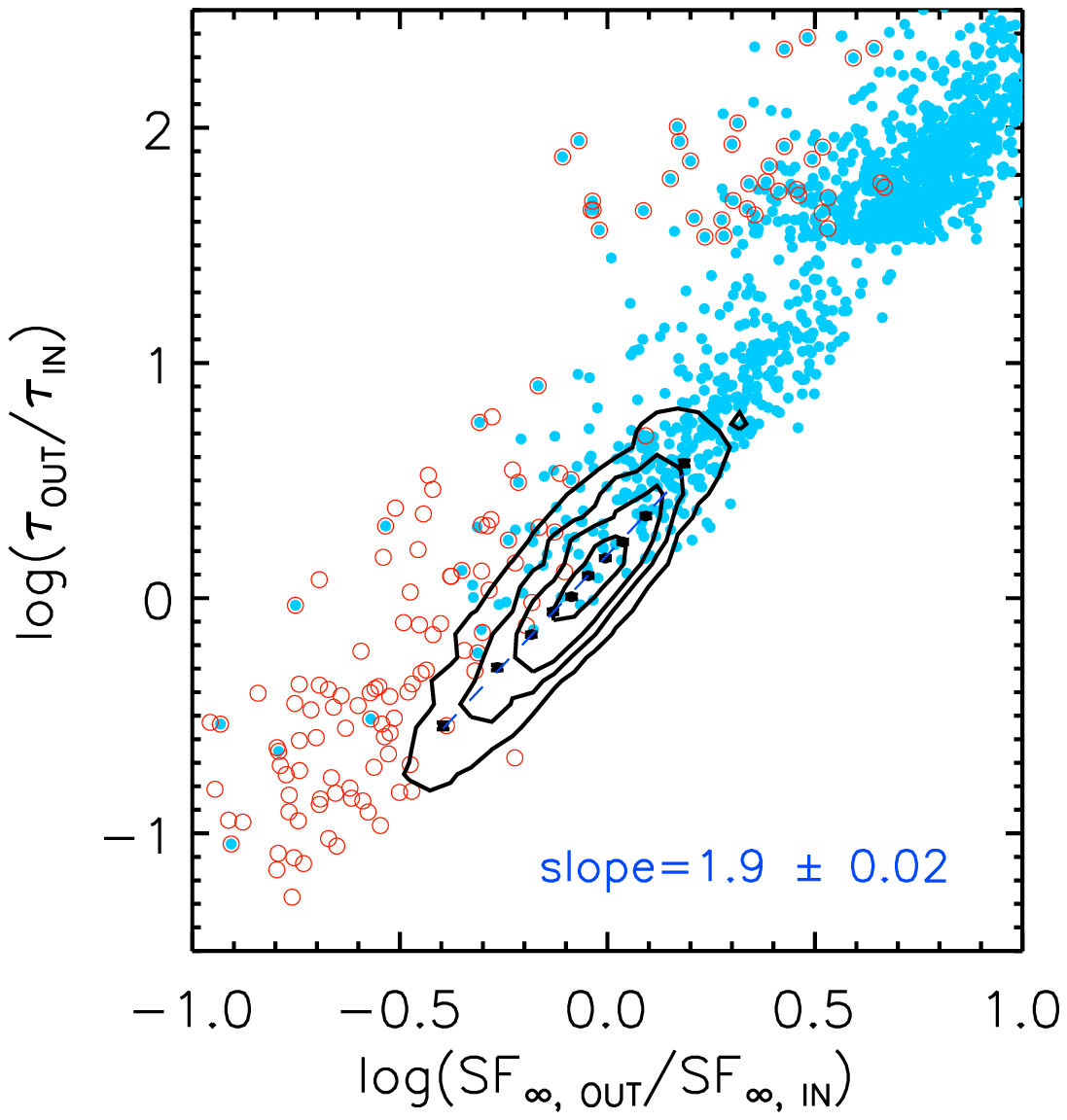}}
\caption{Test of biases in best-fit parameters.  \emph{Left}: The input variability parameters
  $\tau$ and $SF_{\infty}$ were drawn from
  a uniform distribution in $\log \tau$ and $\log SF_{\infty}$ limited
  by the dotted rectangle. 
  The contours show the output distribution after applying the cuts
  described in Section~\ref{sec:cuts}, and the dashed line is a linear
  regression between the output parameters (with the slope 
  listed on the panel). 
  Solid, blue circles show the objects which do not satisfy 
  $\Delta L_{\infty}>0.05$ (these make up 21\% of the starting sample, where 13\% 
  are saturated at $\tau=10^5$ days).  The small black dots show
  the input values for these rejected points. Open, red circles show
  the 3\% that do not satisfy $\Delta L_{noise}>2$. \emph{Right:} 
  Relationship between $\tau_{out}$/$\tau_{in}$ and
  $SF_{out}$/$SF_{in}$. The slope of the cleaned output distribution
  is listed. The sharp edge is due to the saturation limit of $\tau=10^5$ days.
  The contours show the 90\%, 70\%, 50\%, and 20\% levels.}
\label{fig:sftau_fake}
\end{figure*}

Since the correlation between $\tau$ and $SF_{\infty}$  seen in
Figure~\ref{fig:pr_tau_sf} might be expected if the light
curves are not sampled over sufficiently long periods of time 
(see Section~\ref{sec:modelLCs}), we first test whether the
correlation is real or simply an artifact. We generated light curves
as described in Section~\ref{sec:modelLCs} using the time sampling and
photometric uncertainties of the S82 data. 
For each object, the input parameters $\tau_{in}$ and $SF(\infty)_{in}$
are randomly drawn from a uniform distribution in $\log \tau$ and
$\log SF_{\infty}$ limited by the dotted rectangle in the left panel of
Figure~\ref{fig:sftau_fake}.  These artificial light curves are then 
fit to obtain $\tau_{out}$ and $SF(\infty)_{out}$, and the resulting 
distribution is shown by the contours in Figure~\ref{fig:sftau_fake}. 
The open circles show data points that do not satisfy
$\Delta L_{noise}>2$ (3\% of all input points) -- these are concentrated
at small $\tau_{out}$ and $SF(\infty)_{out}$. The closed circles show those that
do not satisfy $\Delta L_{\infty}>0.05$ (21\% of all input points) --
these have been smeared to large values of $\tau_{out}$ and $SF(\infty)_{out}$. 
After omitting points with $\Delta L_{noise}\leq 2$ and $\Delta
L_{\infty}\leq 0.05$, the distribution of the output estimates is
similar to the input distribution and shows no strong correlation
between $\tau_{out}$ and $SF(\infty)_{out}$.  This suggests that the
correlation in Figure~\ref{fig:pr_tau_sf} is largely real, and not an
artifact of sampling and fitting procedures. These results also
justify the selection cuts outlined in Section~\ref{sec:cuts}.  
The right panel of Figure~\ref{fig:sftau_fake} shows the expected
correlation that the larger the overestimate of $\tau$, the larger
the overestimate of $SF_{\infty}$, with a slope between them following
that expected by Eq.~\ref{eq:sfinf} (for $\Delta t << \tau$).
We repeated the test using a uniform input distribution in $\log
\hat{\sigma}$ (the driving amplitude of short-term variations) rather
than in $\log SF_{\infty}$, where
$\hat{\sigma}=SF_{\infty}/\sqrt{\tau}$. In this case, we find the
output $\tau$ and $SF_{\infty}$ are correlated with a power law slope
of $0.18 \pm 0.01$. Since this slope is smaller than that for the
observed distribution in Figure~\ref{fig:pr_tau_sf}, the $\hat{\sigma}$
and $\tau$ must be intrinsically correlated for the S82 sample as well. 

\begin{figure*}[p]
\centering
\includegraphics[width=6in]{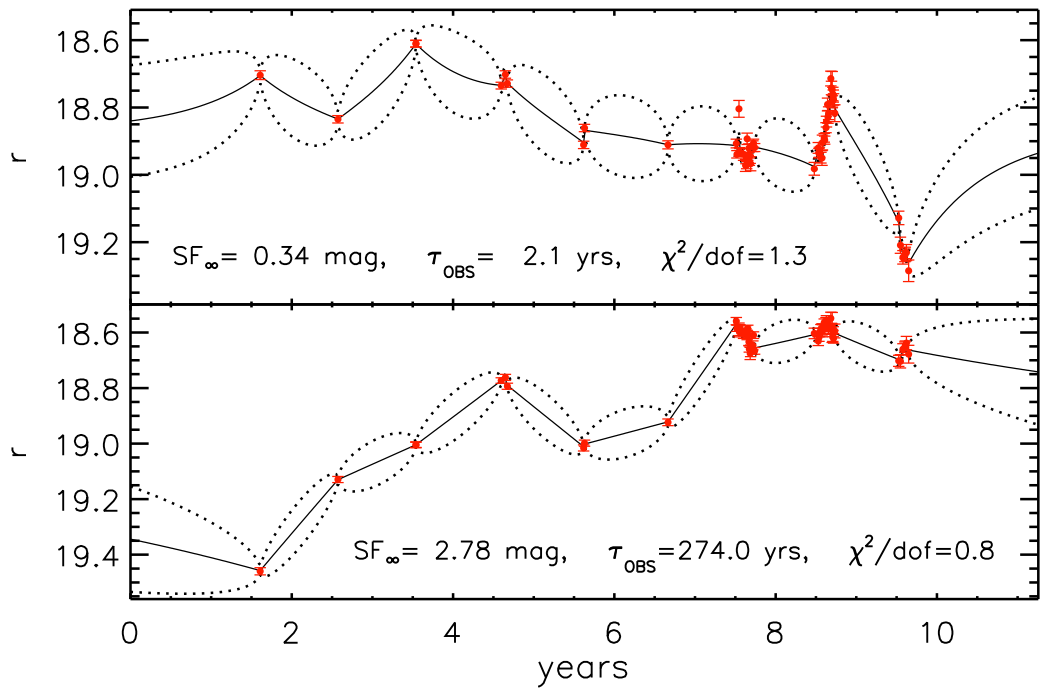}
\caption{\emph{Top panel}: Data points with error bars show the
  observed light curve for a quasar in our final sample, the solid
  line shows the weighted average of all consistent  
  model light curves, and the dotted lines show the $\pm 1\sigma$ range of possible stochastic
  models (see Section~\ref{sec:methodology}). 
  \emph{Bottom panel}: A ``regenerated'' light curve using
  Eq.~\ref{eq:car1sim} and the estimated values of $\tau$ and
  $SF_{\infty}$ listed in the top panel.
  Due to the poor time sampling of the light curve, the best fit to
  the ``regenerated'' light curve has a run-away time scale (listed at
  the bottom), with $\Delta L_{\infty}=0.003$.}
\label{fig:LCeg}
\end{figure*}

\begin{figure*}[p]
\centering
\includegraphics[width=6in]{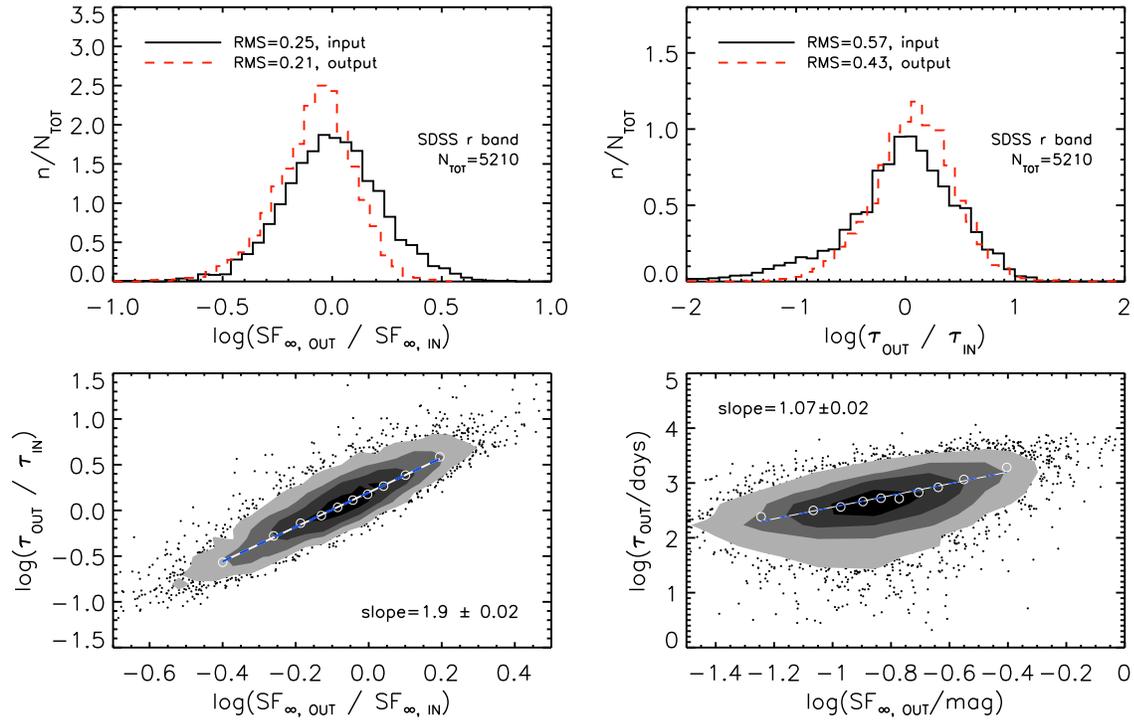}
\caption{Results from ``regenerating'' light curves derived from the 
  observed $\tau$ and $SF_{\infty}$ (see Section~\ref{sec:tests}). 
  The solid, black lines in the \emph{top panels} show the distribution of input  $SF_{\infty}$ (\emph{left})
  and $\tau$ (\emph{right}) normalized by the median input value, 
  and the ratio of output to input values are
  shown by the red, dashed lines. The
  \emph{bottom panels} show the scatter in the output-to-input
  ratios (\emph{left}), and the relationship between the output parameters 
  (\emph{right}). Listed in each bottom panel is the slope of a linear fit
  (dashed line) to the median values (open circles).  Contours show
  the 90\%, 70\%, 50\%, and 20\% levels. }
\label{fig:stotest}
\end{figure*}

For our second test, we used the best-fit $\tau$, $SF_{\infty}$, and
$\mu$ for the S82 sample to generate new light curves with the same
time sampling and photometric uncertainties as the S82 data. 
By comparing the output and input parameter distributions, 
we can estimate how much the intrinsic stochasticity and
the time sampling issues affect the results. 
Figure~\ref{fig:LCeg} compares an observed and
``regenerated'' light curve, where the differences are due to the
stochastic nature of the process. The fit parameters for the
``regenerated'' light curve can be very different because of how the
particular realization is affected by the time sampling, as
illustrated in Figure~\ref{fig:LCeg}.
Figure~\ref{fig:stotest} shows the ratio of output to input
distributions for both $\tau$ and $SF_{\infty}$.  The input
distributions normalized by their median values are also shown to
illustrate their dynamic range. These
two distributions are compared to each other in order to estimate the
effect of fitting errors (the ratio of output to input should
be a delta function centered at 1 for perfect time sampling). 
The bottom-right panel shows that the correlation between $\tau$ and
$SF_{\infty}$ becomes slightly weaker than that seen in 
Figure~\ref{fig:pr_tau_sf}.
Based on these results, we conclude that the uncertainties due to sparse
sampling and limited lengths of the S82 light curves can account for
71\% of the spread in $SF_{\infty}$ and 57\% of that in $\tau$.  
As shown in Section~\ref{sec:modelLCs}, very long light curve lengths
are needed to estimate accurate time scales and asymptotic amplitudes. 
Nevertheless, the observed distributions indicate that the 
underlying intrinsic distributions of $\tau$ and $SF_{\infty}$ have
finite widths that are similar to the observed widths.

\subsection{Relationship between the Individual SFs and the Ensemble SF}
\label{sec:ringberg}

The distribution of $\tau$ is important to consider when
relating the ensemble SF, such as those determined using
two-epoch datasets,  to the SFs for individual light curves. 
The analysis based on two-epoch data measures {\it the mean value} of
the SF, but provides {\it no information} about the SF variance among individual 
objects. To measure the latter, individual light curves must be available.

\begin{figure*}[p]
\centering
\includegraphics[width=6in]{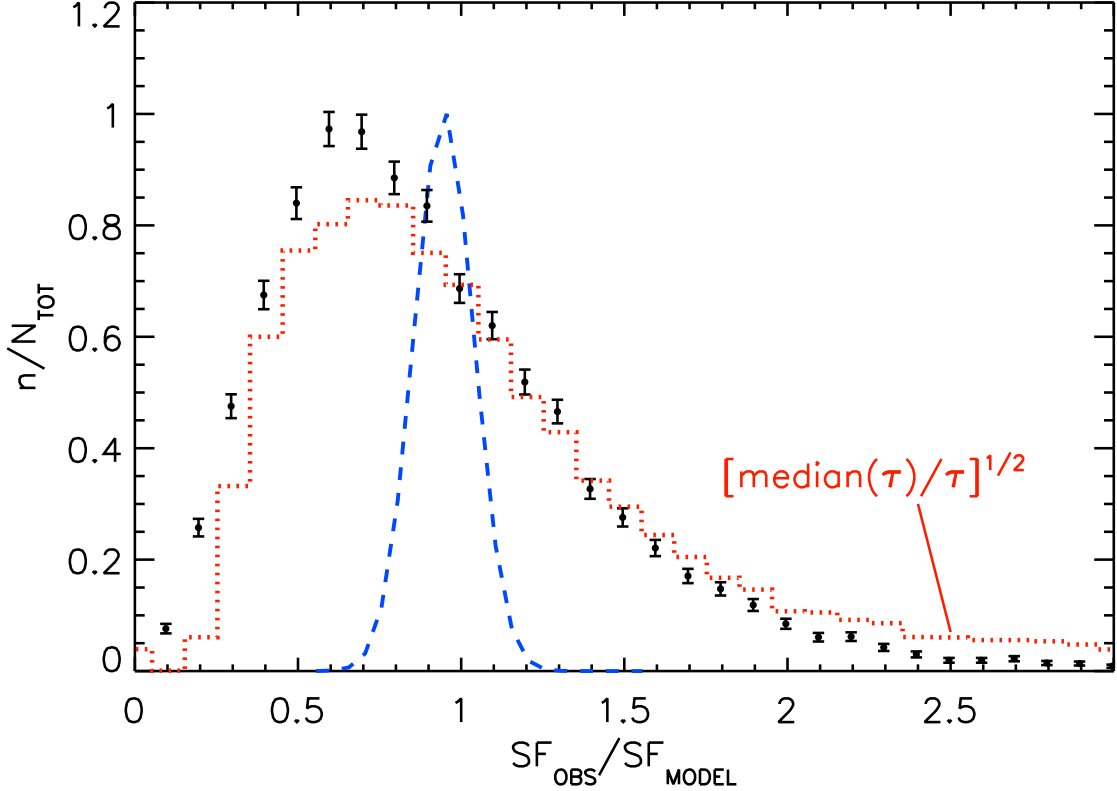}
\caption{The symbols with error bars show the 
  distribution of $SF_{obs}/SF_{model}$ from Figure 2 in
  MacLeod et al.~(2008), where $SF_{obs}$ is the observed structure
  function for individual S82 light curves for a fixed observed-frame time
  lag of 1 year, and $SF_{model}=1(1+0.03M_i)(\Delta
  t_{RF}/\lambda_{RF})^{0.47}$. The distribution from the ensemble analysis by 
  I04 (based on two-epoch SDSS data) is shown by the dashed curve, scaled by a factor of 0.2. 
  The dotted histogram is the distribution of $\sqrt{\rm
  median(\tau)/\tau}$ from this work, where $\tau$ is the
  characteristic time scale for each light curve. Its agreement with the
  $SF_{obs}/SF_{model}$ distribution for individual light curves shows
  that most of the scatter pointed out by MacLeod et al.~(2008) is due to a
  finite width of the $\tau$ distribution. }
\label{fig:ringberg}
\end{figure*}

MacLeod et al.~(2008) analyzed individual light curves for S82 quasars in 
order to test the common assumption that photometric observations at two 
epochs for a large number of quasars will reveal the same statistical properties 
seen in light curves for individual objects. They found that the dependence of the 
mean SF computed using SFs for individual light curves on luminosity, rest-frame wavelength and 
time lag is indeed qualitatively and quantitatively similar to that derived from two-epoch 
observations of a much larger sample. However, they also found that the scatter in the 
light-curve based SFs  for fixed values of $M_i$, $\lambda_{RF}$ and $\Delta t_{RF}$
is very large, and in fact, similar to the scatter for the whole sample (see Figure~\ref{fig:ringberg}).  
This large scatter was attributed to an intrinsic stochasticity of unknown origin. The new 
model-based analysis discussed here allows us to explain this puzzling result as a 
consequence of the finite width of the $\tau$ distribution.

An important piece of information missing from the MacLeod et al.~(2008) analysis
is the existence of a characteristic time scale $\tau$. In MacLeod et al.~(2008), the 
individual structure functions ($SF$) were computed following the standard approach
with a fixed observer's frame time lag of 1 year (and with rest-frame
time lags spanning 100 to 300 days). However, Eq.~\ref{eq:sfinf}
indicates that the SF at small time lags should be proportional 
to $1/\sqrt{\tau}$, indicating that the observed structure functions will vary
between quasars even if they have similar $SF_{\infty}$. Indeed, 
the distribution of $\sqrt{\rm median(\tau)/\tau}$ has an almost
identical shape and width as that of $SF/SF_{model}$ (see
Figure~\ref{fig:ringberg}). We therefore conclude that variations in
$\tau$ are responsible for most of the scatter in $SF(\Delta t <<
\tau)$ for quasars with similar luminosity, rest wavelength, and time lag.
In addition, it is likely that the $\tau$ distribution is responsible for the exponential 
tails of the magnitude difference distribution for quasars reported by I04 and
Sesar et al.~(2006; see MacLeod et al.\ 2010, in prep., for further discussion). 
Therefore, the published structure function results based on two-epoch
datasets can only be interpreted in the context of Eq.~\ref{eq:ensSF}. 

\subsection{Dependence on the Underlying PSD}
\label{sec:PSD}
The analysis throughout this paper assumes that the variability is
described by a damped random walk, which has a power spectral
distribution (PSD) described by $PSD \propto f^{-2}$ at frequencies
$f>(2\pi\tau)^{-1}$, flattening to a constant at lower frequencies. In
this Section, we investigate the sensitivity of our resulting
parameter distributions to the possibility that the low frequency part
of the PSD is not flat. For example, the X-ray variability of Seyfert
galaxies is well-described by a broken power law with a slope of $-2$
at high frequencies, breaking to a shallower slope ($-1$) at low
frequencies (Ar{\'e}valo et al.~2008a). Since the optical and X-ray
variations are correlated (Ar{\'e}valo et al.~2008b), it is plausible
that the optical variability might have a similar underlying PSD. On
the other hand, as reviewed by  McHardy (2010), the optical and X-ray
fluctuations are only correlated on time scales which are shorter than
the typical characteristic optical time scale (i.e., on time scales
corresponding to the $1/f^2$ part of the optical power spectra). If
the damped random walk is a good description for the optical
variability, then we might expect that the optical and X-ray
fluctuations are no longer correlated on time scales longer than
$\tau$, as the optical fluctuations resemble white noise on these time
scales. 

\begin{figure*}[p]
\centering
\includegraphics[width=6in]{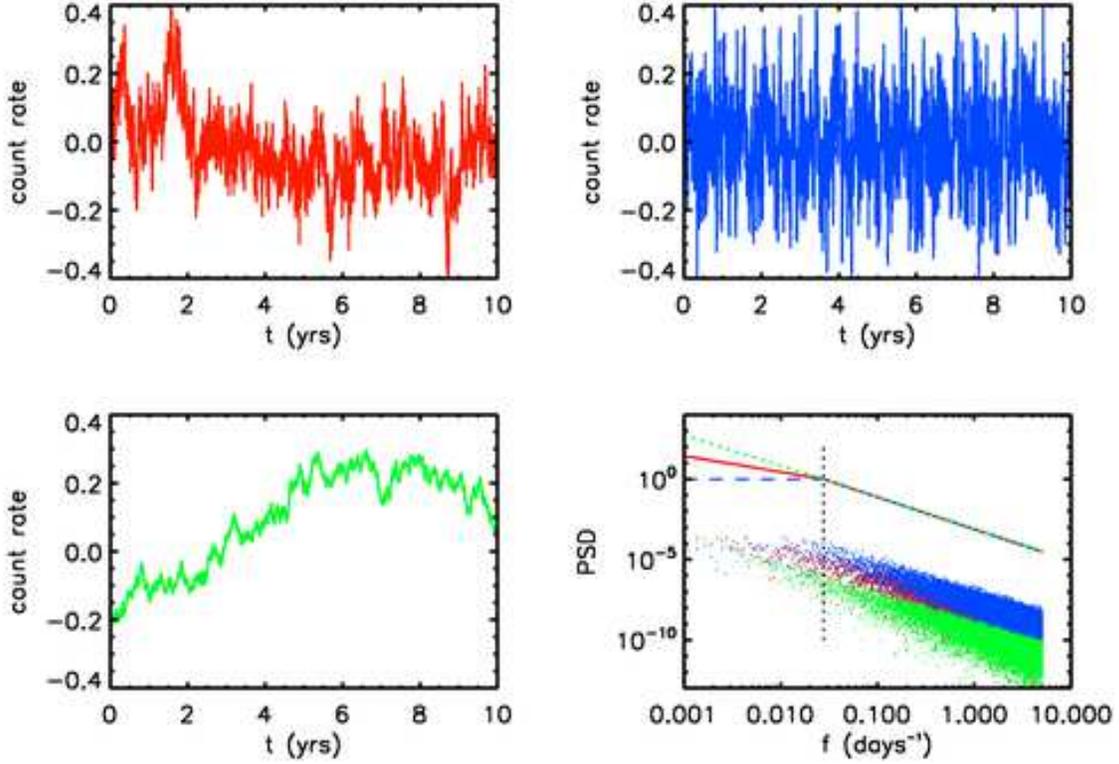}
\caption{ 
  \emph{First three panels}: Simulated light curves using $\tau=5.8$ days,
  $rms=0.124$ mag, and a PSD described by  $PSD \propto f^{\alpha}$, with
  $\alpha=-2$ for $f>(2\pi\tau)^{-1}$, and $\alpha=-1$ (\emph{red;
  top-left}), $\alpha=0$ (\emph{blue; top-right}), or $\alpha=-1.9$
  (\emph{green; bottom-left}) for $f<(2\pi\tau)^{-1}$. The lines in
  the \emph{last panel} show each PSD from which the light curves are
  generated, with the \emph{black dotted line} indicating the break
  frequency. As a check, the PSD was then computed for the simulated
  light curves, shown with \emph{colored dots}. The y-axis in the
  bottom-right panel is in arbitrary units.  
}
\label{fig:simLC_psd}
\end{figure*}

We consider three cases for a broken, or bending, PSD described by 
$PSD \propto f^{-2}$ at high frequencies and $PSD \propto f^{\alpha}$
at low frequencies. The first case is where $\alpha=0$, which is a
damped random walk. The second case is where $\alpha=-1$, and the
third is where $\alpha=-1.9$, which is nearly a constant power-law
slope. In each case, $\sim$7000 light curves are simulated using the
algorithm from Timmer \& Koenig (1995) with the chosen PSD. For each
realization, the break frequency is set to $(2\pi\tau)^{-1}$ and the
total $rms$ is fixed, using the observed $r$-band $\tau$ and $rms$
values for the S82 quasars that satisfy the selection criteria in
Section~\ref{sec:cuts}. Therefore, the observed distributions in
Figure~\ref{fig:pardist} (for the $r$-band) are the ``input''
values. Figure~\ref{fig:simLC_psd} shows 3 example light curves
simulated using $\alpha=-1$ (red), $\alpha=0$ (blue), or $\alpha=-1.9$
(green). The lines in the last panel show each PSD from which the
light curves are generated, with the black dotted line indicating the
break frequency. As a check, the PSD was computed for each simulated
light curve, and as seen by the colored dots, the shapes match the
input.   

\begin{figure*}[p]
\centering
\includegraphics[width=6in]{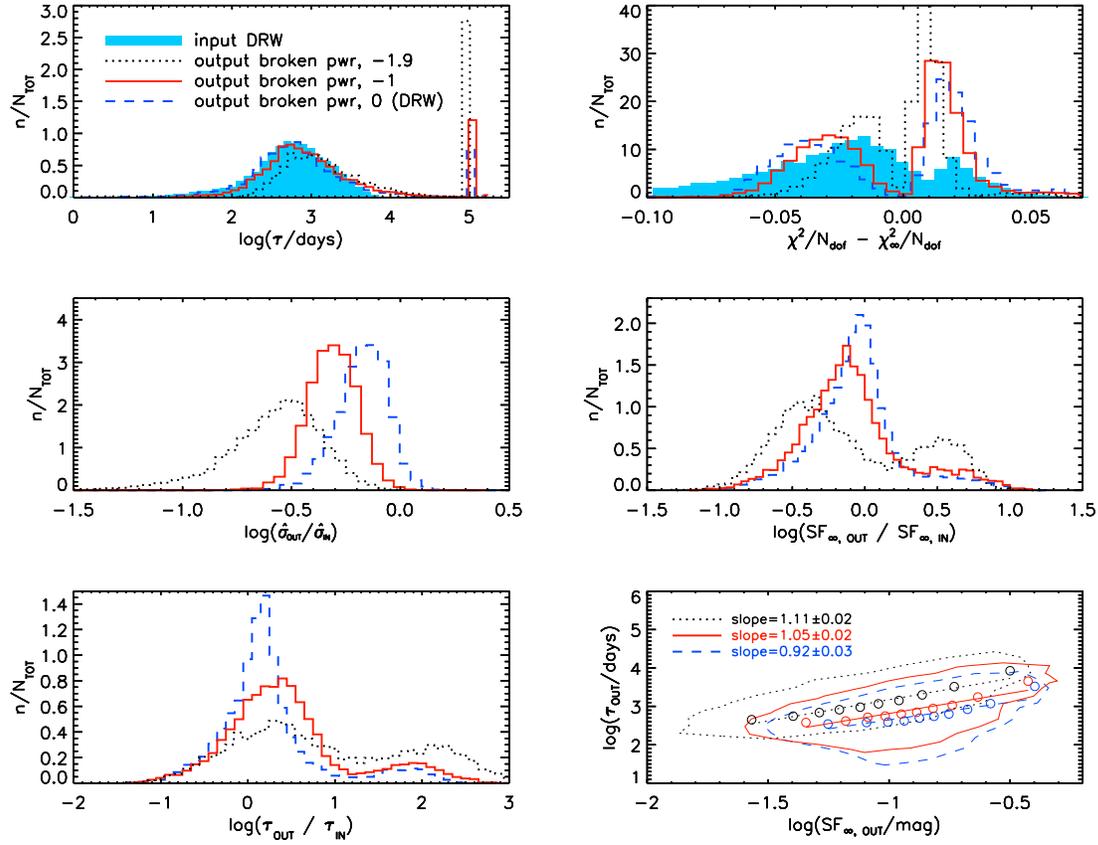}
\caption{\footnotesize In the top panels, the \emph{filled histogram} shows the
  (``input'') distributions of $\tau$ and 
  $\Delta \chi^2_{\infty} = \chi^2/N_{dof} - \chi^2_{\infty}/N_{dof}$   
  for $r$-band S82 light curves ($\sim$7000 total). 
  This distribution of $\tau$, along with the $r$-band $rms$ values, are
  used to generate $\sim$7000 realizations of noise processes with
  $\alpha=-2$ for $f>(2\pi\tau)^{-1}$ and $\alpha=0$, $\alpha=-1$, or
  $\alpha=-1.9$ for $f<(2\pi\tau)^{-1}$.  The \emph{dashed blue},
  \emph{solid red}, and \emph{dotted} lines show the best-fit
  (``output'') distributions when modeling these three processes,
  respectively, as a damped random walk (DRW). Here, the simulated
  observations are spaced every 5 days over 10 years with typical
  errors of 0.01 mag.  The output $\hat{\sigma}$, $SF_{\infty}$, and
  $\tau$ are compared to the input values in the next three panels. In
  the last panel, the correlation between the output $log(SF_{\infty})$ and
  $log(\tau)$ is shown using a linear fit to the median values
  (\emph{open circles}), with the slopes listed in the legend for each
  case. Contours show regions containing 90\% of the data points. 
}
\label{fig:wellsamp}
\end{figure*}

\begin{figure*}[p]
\centering
\includegraphics[width=6in]{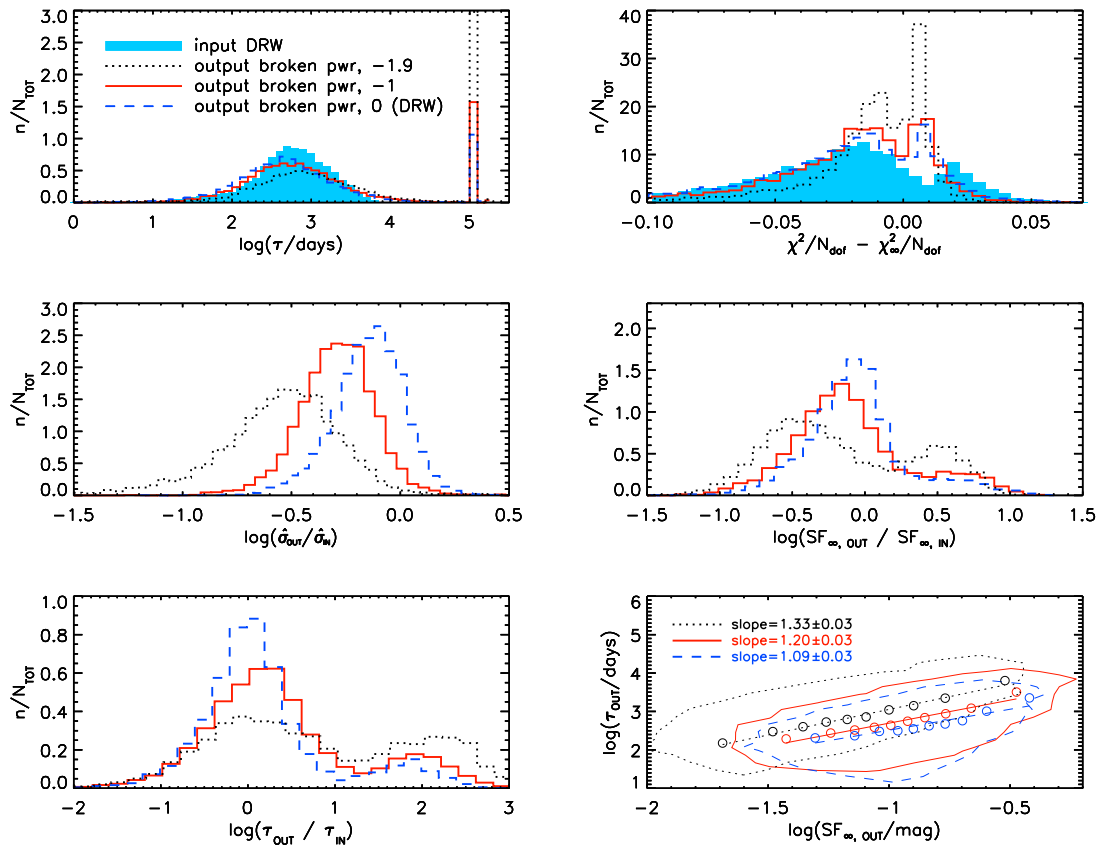}
\caption{ As in Figure~\ref{fig:wellsamp}, except with the Stripe 82
  time sampling and photometric accuracy imposed on the simulated
  light curves. 
}
\label{fig:S82samp}
\end{figure*}

Each light curve was simulated over 100 years and then truncated to
the 30--40 year segment in order to account for the additional
variability and bias in the inferred $SF_{\infty}$, which may result
from a red noise leak or an incorrect specification of the PSD model  
(see Uttley et al.~2002).  
The simulated light curves were then modeled as a damped random walk
to obtain the (``output'') parameter distributions shown in
Figures~\ref{fig:wellsamp} and~\ref{fig:S82samp}. For
Figure~\ref{fig:wellsamp}, the simulated observations are spaced every
5 days over 10 years with typical errors of 0.01 mag, and for
Figure~\ref{fig:S82samp}, the S82 window function is imposed (i.e.,
all light curves have the S82 time sampling and photometric accuracy).
The filled histograms show the input distributions and are the same
for both Figures.

It is clear from comparing the input and output parameter
distributions that $\tau$ and $SF_{\infty}$ map simply onto the
values of the power-spectral break time scale and amplitude. Indeed,
the correlation between $\tau$ and $SF_{\infty}$ is preserved in the
output distributions, as seen in the bottom-right panels of
Figures~\ref{fig:wellsamp} and~\ref{fig:S82samp}.
However, as seen in the top- and bottom-left panels, as the PSD slope at
low $f$ steepens to $\alpha=-1.9$, the number of ``run-away'' time
scales (where $\tau$ saturates at $10^5$ days) increase due to that
fact that $\tau$ can no longer be constrained. 
Whereas for the S82 data, $\sim$20\% of light curves were rejected as
run-away cases (see step 3 of Section~\ref{sec:cuts}), for an
$\alpha=-1.9$ PSD, the run-away fraction is 30\%. This significant
increase rules out the $\alpha=-1.9$ PSD as the correct model, as we
do not see this large  run-away fraction in the data. The fractions
are similar for the $\alpha=-1$ and $\alpha=0$ cases, suggesting that
the correct model has $-1<\alpha<0$. 
The increasing fraction of run-away $\tau$  with decreasing $\alpha$ allows one to
distinguish between each PSD in the well-sampled case, as seen in the
top-right panel of Figure~\ref{fig:wellsamp}. In this panel, the
difference between the $\chi^2_{pdf}$ for the best-fit damped random
walk and that for a $\tau \rightarrow \infty$ solution, $\Delta
\chi^2_{\infty} = \chi^2/N_{dof} - \chi^2_{\infty}/N_{dof}$, is shown
for each input PSD. The total distribution shows two peaks; that on
the left corresponds to cases where the model is able to constrain
$\tau$.  For well-sampled light curves (Figure~\ref{fig:wellsamp}), 
the case with $\alpha=0$ (dashed line) yields an overall lower $\Delta
\chi^2_{\infty}$ and therefore a more likely damped random walk fit,
as expected. However, for the S82 sampling, it is difficult to
distinguish between the red and blue lines ($\alpha=0$ and $-1$), and
the (input) observed distribution is similar to both. Therefore,
within the limited S82 sampling, we are unable to distinguish reliably
between a damped random walk and a $1/f$ PSD on long time scales.

\section{Dependence of Variability Parameters on Luminosity, Wavelength,
  Redshift, and Black Hole Mass} 
\label{sec:trends}

\begin{figure*}[p]\centerline{
    \includegraphics[width=2.5in]{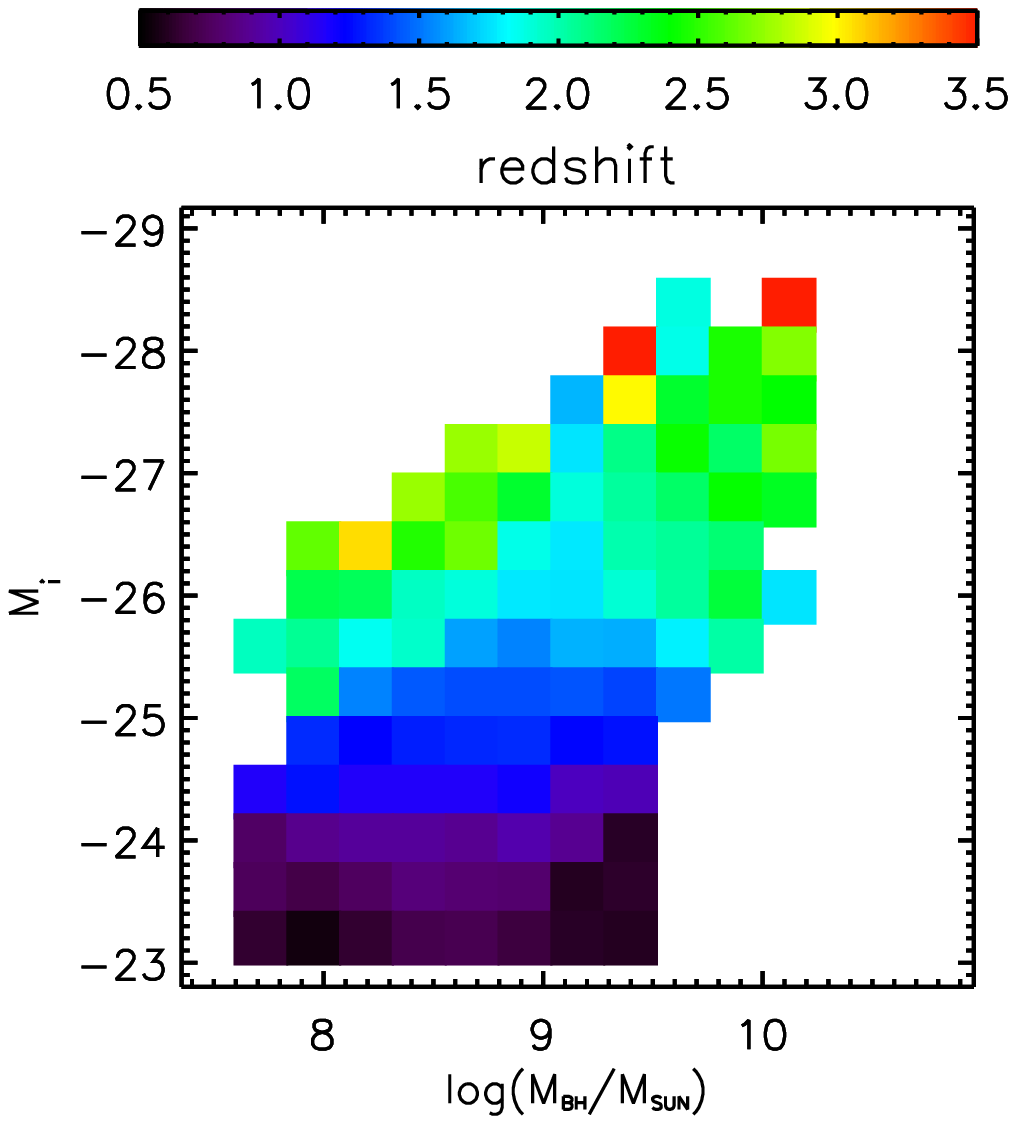}
    \includegraphics[width=2.5in]{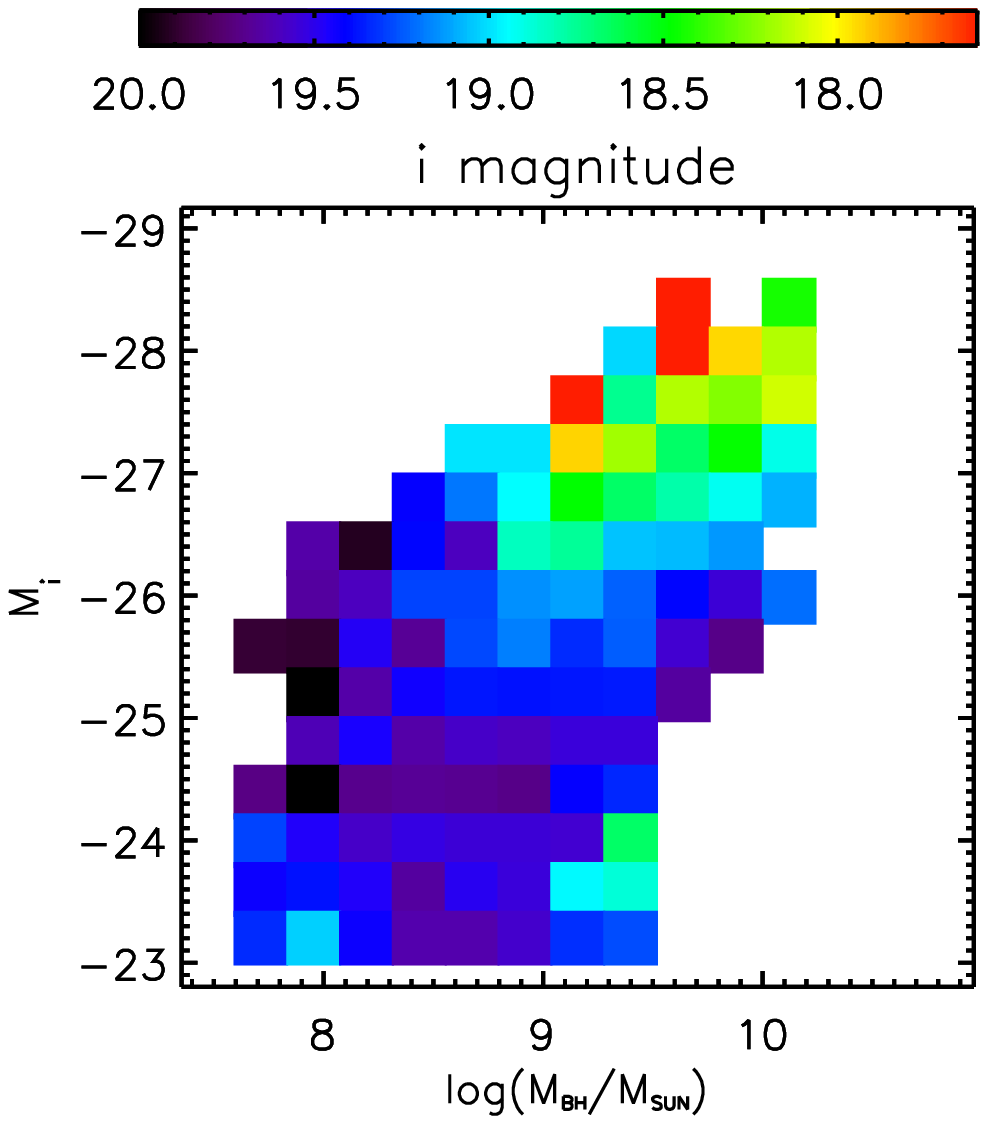}}\centerline{
    \includegraphics[width=2.5in]{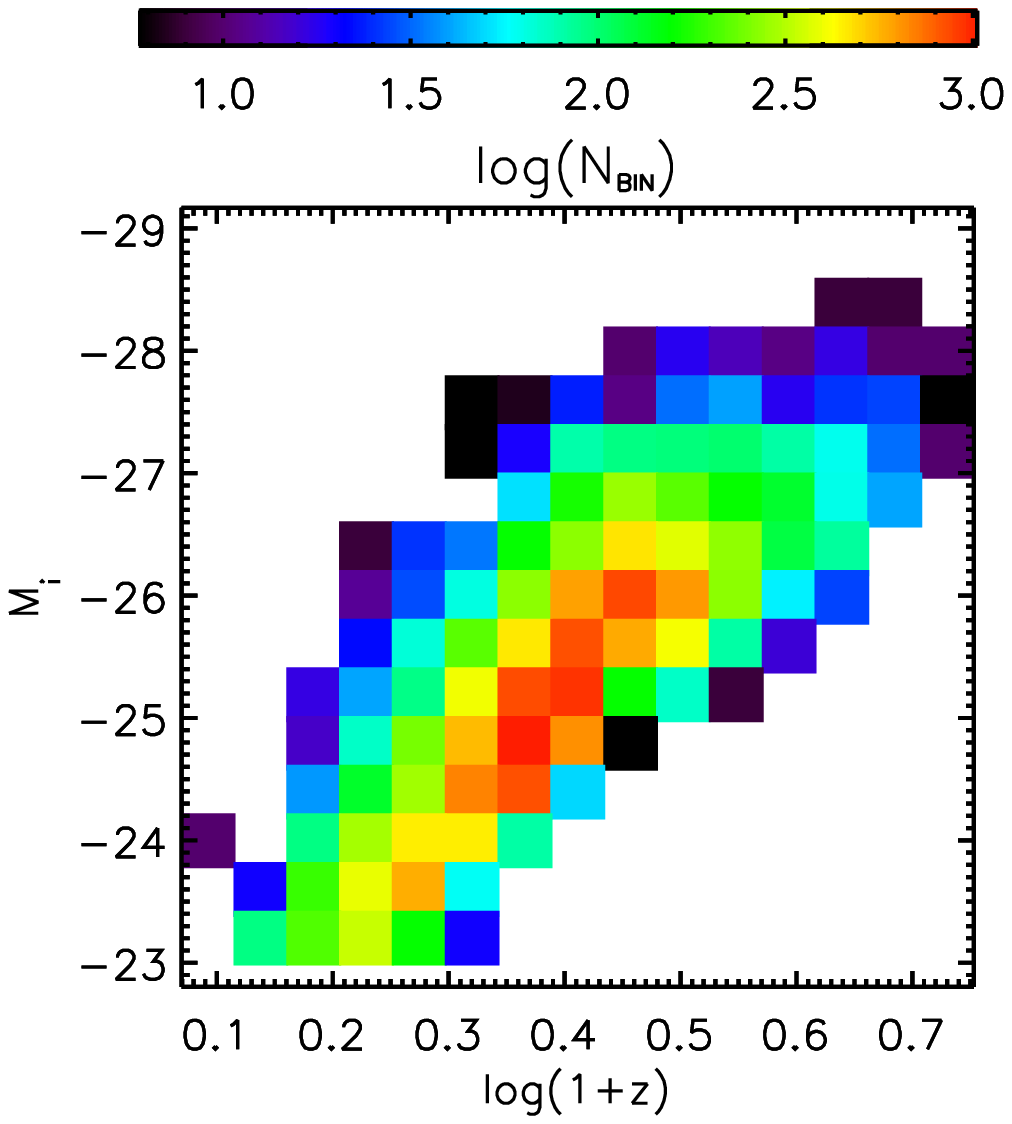}
    \includegraphics[width=2.5in]{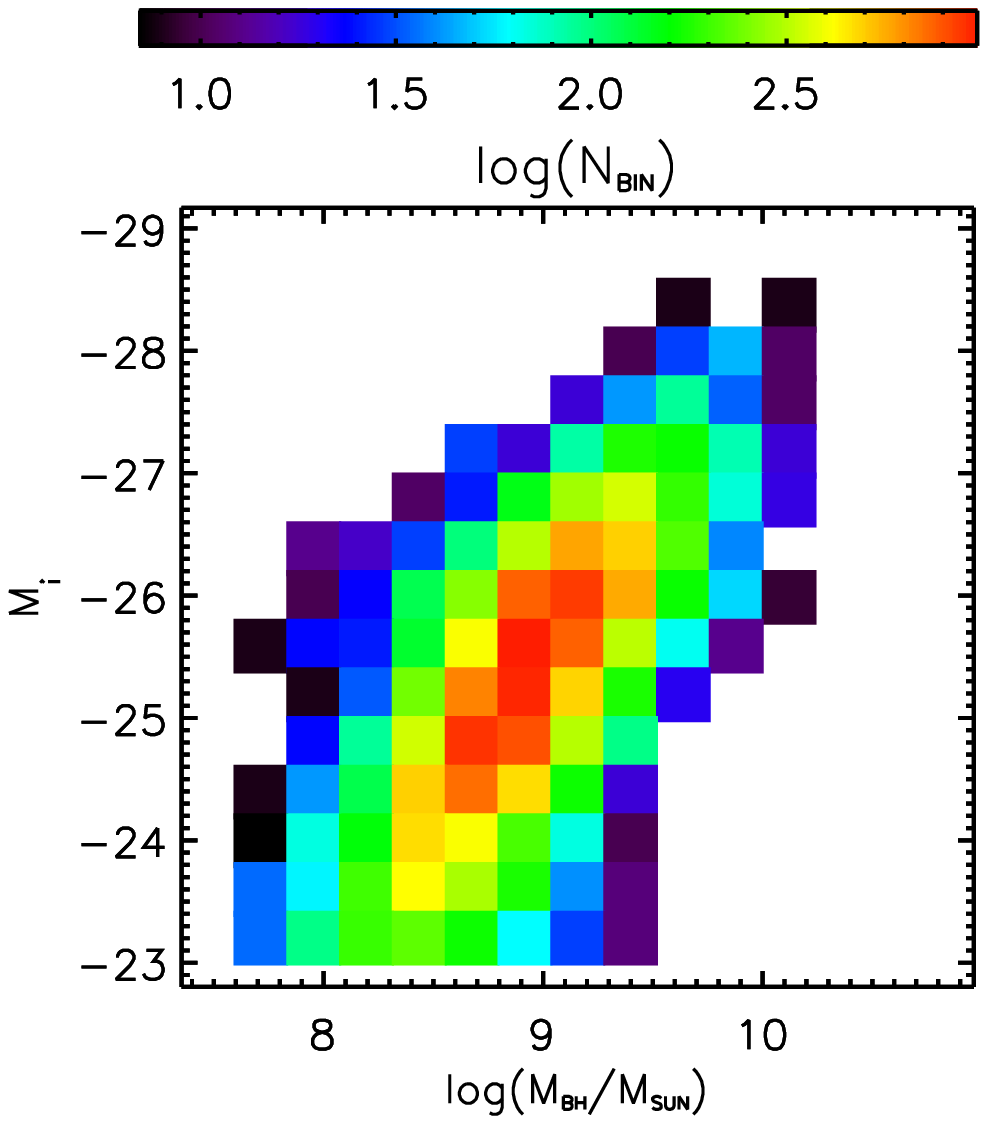}}
\caption{The \emph{top panels} show the distribution of quasars in
  redshift (\emph{left}) and i magnitude (\emph{right}) on a grid of
  $M_{BH}$ vs. $M_i$. The \emph{bottom panels} show the number of data
  points in bins of $M_{BH}$ (or redshift) vs. $M_i$.}
\label{fig:hiddensys}
\end{figure*}

Next, we discuss correlations between each of the variability
parameters, $\tau$ and $SF_{\infty}$, 
and the four available physical parameters: rest-frame wavelength
($\lambda_{RF}$), redshift ($z$), absolute i-band magnitude ($M_i$), and 
black hole mass ($M_{BH}$). It is important to fit a multiple
regression to each of the physical parameters, functions of the form
$\tau(\lambda_{RF},M_i,M_{BH},z)$, because of the correlations between
physical parameters. 
For example, when searching for a correlation between $SF_{\infty}$ and 
$M_{BH}$, one must take into account that a more luminous quasar hosts
a more massive black hole (e.g., Kollmeier et al.~2006), or else 
a trend with luminosity may be mistaken for a trend with black hole mass. 
A similar example is the well-known luminosity-redshift ($L$--$z$)
degeneracy seen in flux-limited samples of quasars.  
Magnitude limits result in the illusion that only the most luminous
quasars are seen at high $z$, and therefore the 
observed $L$ increases with $z$ independent of reality. 
Having a large sample size helps to alleviate these degeneracies. 
For example, one can look for trends using 2-dimensional
grids in any two physical parameters of interest. 
The numbers of data points per two-dimensional bin of $M_i$ and redshift, and similarly 
for $M_i$ and $M_{BH}$, are shown in the bottom panels of Figure~\ref{fig:hiddensys}.  
The two left panels of Figure~\ref{fig:hiddensys} show the selection
effect that quasars at higher redshift must have higher 
luminosities to be included in the survey. From the two-dimensional
distribution in the top-left panel, we can see that this $L$--$z$
degeneracy is essentially independent of $M_{BH}$. Furthermore, 
shorter rest-frame wavelengths are probed at higher redshifts for two reasons:
 first, quasars emitting at shorter rest wavelengths must be at higher redshifts  
in order to be observed within the $ugriz$ filters, and second, quasars at high 
redshifts are closer to their Eddington limit as a result of the
$L$--$z$ degeneracy, and possibly cosmological downsizing 
(e.g., Kollmeier et al.~2006). Therefore, any dependence of
variability on wavelength must be accounted for when considering a  
dependence with redshift (or luminosity). 

When fitting power laws to a large number of data points throughout this paper, we fit
to the median values in each bin of the independent variable, where
all bins have the same number of data points $N$.  The errors in the
medians are computed as 0.93(IQR)$/\sqrt{N-1}$, where IQR is the 25\%
-- 75\% interquartile range, and these constrain our formal errors 
in the power law slopes.

\subsection{Trends with Rest-frame Wavelength}
\label{sec:trendslRF}

We start by examining the wavelength dependence of the variability parameters. 
Since there are multiple bands for each quasar, this dependence can be
determined for individual sources for which $z$, $M_i$, and $M_{BH}$ are fixed. 
The rest-frame wavelength is found by dividing the observed bandpasses 
($3520$, $4800$, $6250$, $7690$, and $9110$ \AA\ for $ugriz$, respectively) 
by $(1+z)$. We fit a power law $f\propto (\lambda_{RF}/4000\AA )^B$ to 
the estimates of $\tau$ and $SF_{\infty}$ for every quasar observed in at least
two filters ($\sim$8,000 quasars). The median values are $B=0.17\pm
0.02$ and $-0.479\pm 0.005$ for $\tau$ and $SF_{\infty}$,
respectively. We searched for significant correlations between
$B$ and other physical parameters and did not find any. We use these
median values to correct for the wavelength dependence of $\tau$ and
$SF_{\infty}$ before searching for their correlations with other
physical parameters in Sections~\ref{sec:trendsmedian},~\ref{sec:Edd},
and~\ref{sec:radio}.  

\begin{figure*}[p]
   \centerline{
\includegraphics[width=6in]{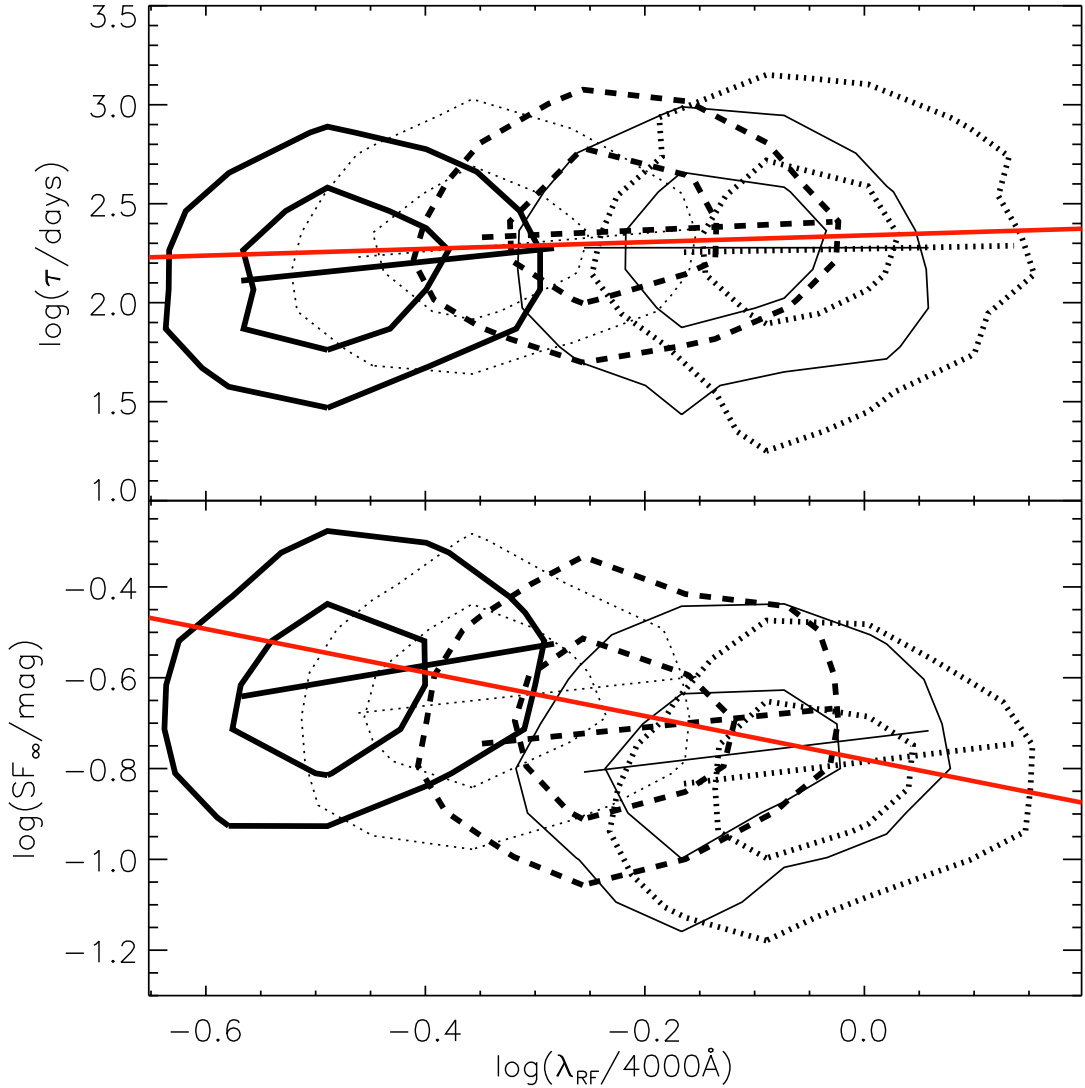}}
\caption{Distribution of the rest-frame time scale
  $\tau$ (\emph{top panel}) and long-term structure function
  $SF_{\infty}$ (\emph{bottom panel}) as a function of rest-frame wavelength
  $\lambda_{RF}$. The different contours show the 70\% and 30\%
  levels for the $u$, $g$, $r$, $i$, and $z$ (from left to right)
  bands. The best-fit power law  for each subsample is shown with the
  same line style as the contours. The thick red line connecting the
  left and right axes shows the median power law slope derived by
  fitting individual quasars (see text), and has a 
  value of $B=0.17$ and $-0.479$ for $\tau$ and $SF_{\infty}$, respectively.
  The slope within each band from fitting ensembles of quasars can be
  quite different from the overall slope  because of the $L$--$z$
  degeneracy (see text). 
  }
\label{fig:pr_lRF}
\end{figure*}

This method of fixing $B$ before investigating other correlations naturally eliminates any 
degeneracies between rest wavelength and the other physical parameters. This is 
especially important in the case of the variability amplitude, $SF_{\infty}$.
Figure~\ref{fig:pr_lRF} shows how the two variability parameters vary with
$\lambda_{RF}$ in each of the $ugriz$ filters. The median power law
from the above analysis, shown as a straight line stretching across
all wavelengths, accurately traces the overall trend.  However, when
the data are fit to ensembles of quasars in each band separately, the
slopes for $SF_{\infty}$, shown by the short lines for each filter,
are very different ($\sim$0.4). This difference is a consequence of
the correlation between $L$ and $z$: for a given band, shorter
$\lambda_{RF}$ corresponds to higher $z$, but at higher $z$, quasars
have higher $L$  and thus smaller $SF_{\infty}$ (see below), creating
a bias in the inferred slope of the wavelength dependence.

\subsection{Trends with Luminosity, Redshift, and Black Hole Mass}
\label{sec:trendsmedian}
\begin{figure*}[p]
   \centerline{
     \includegraphics[width=2.5in]{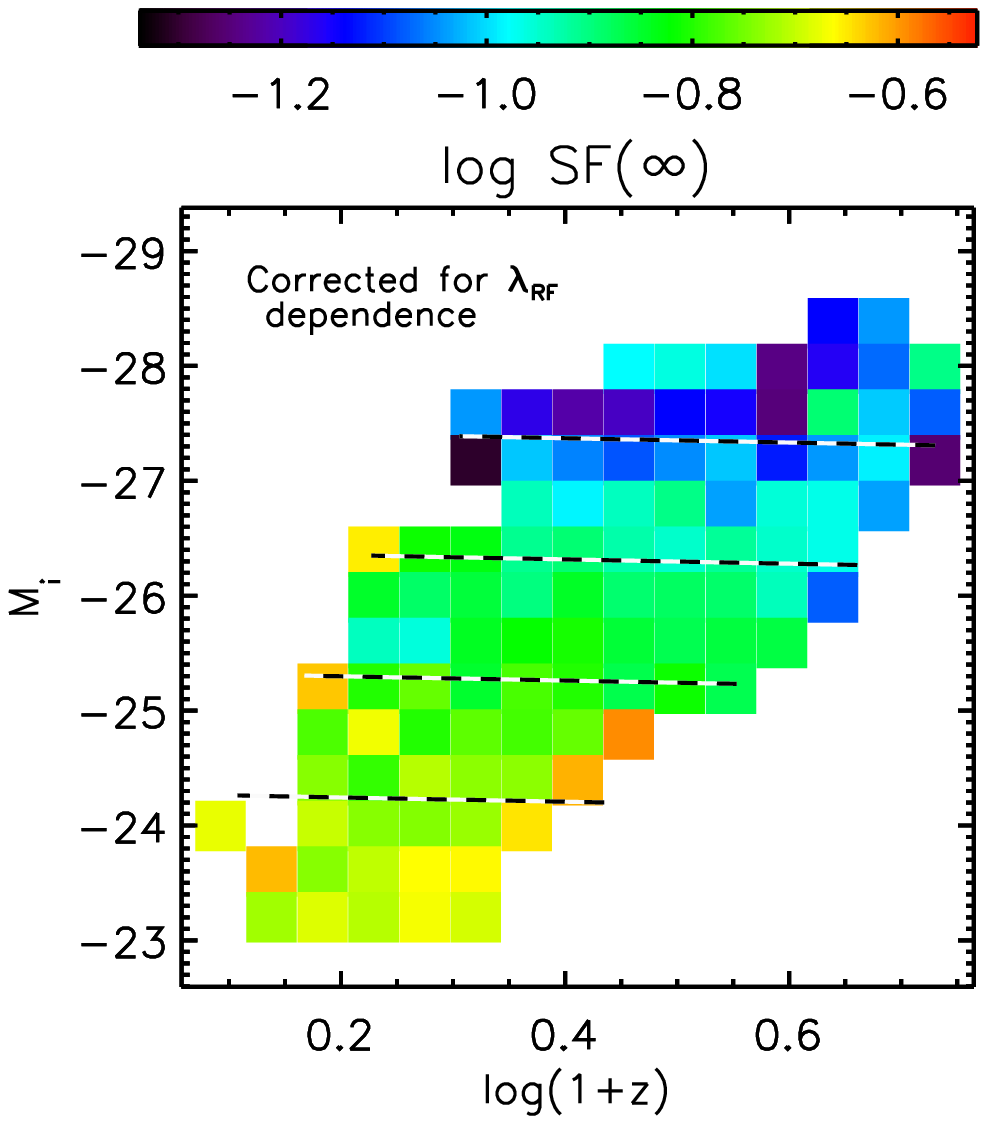}
     \includegraphics[width=2.5in]{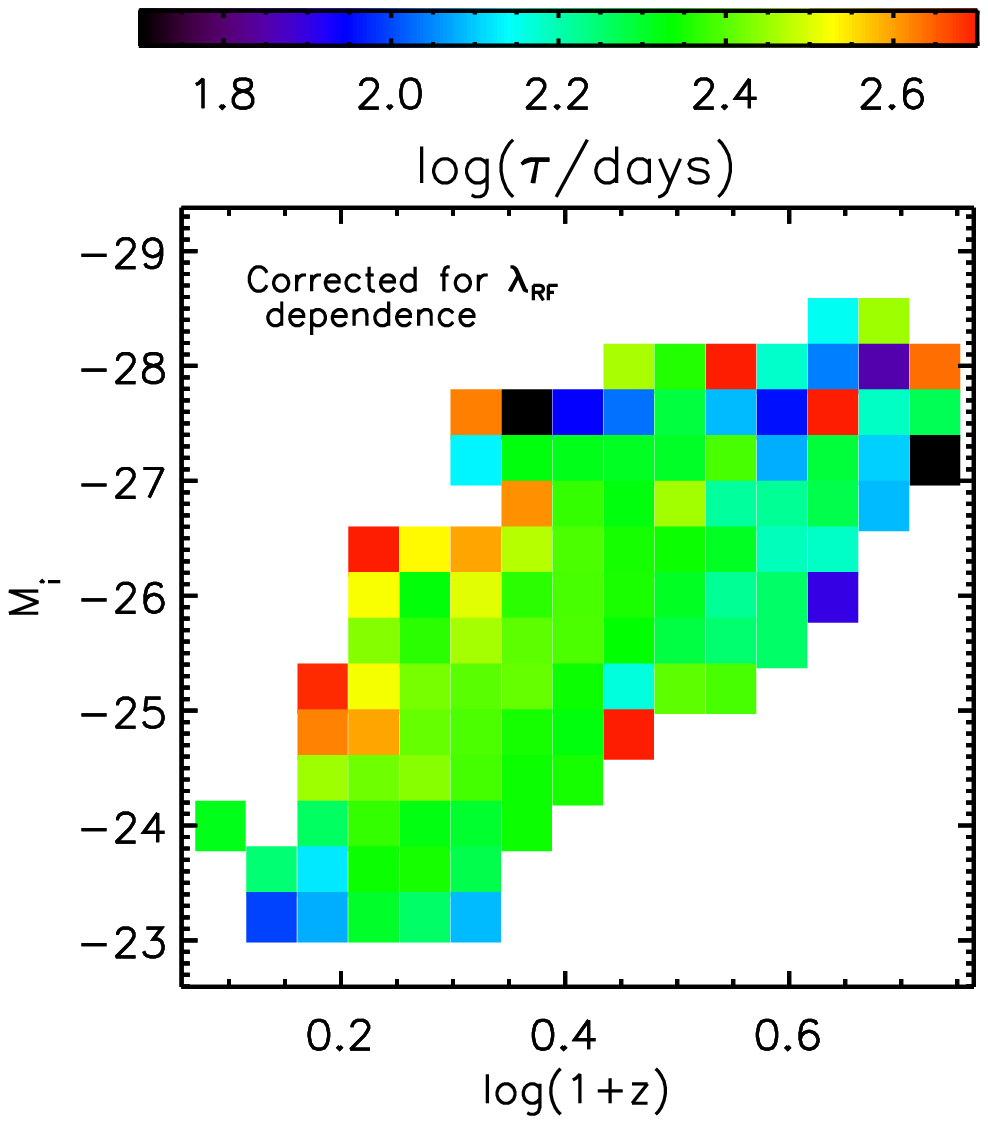}}\centerline{
     \includegraphics[width=2.5in]{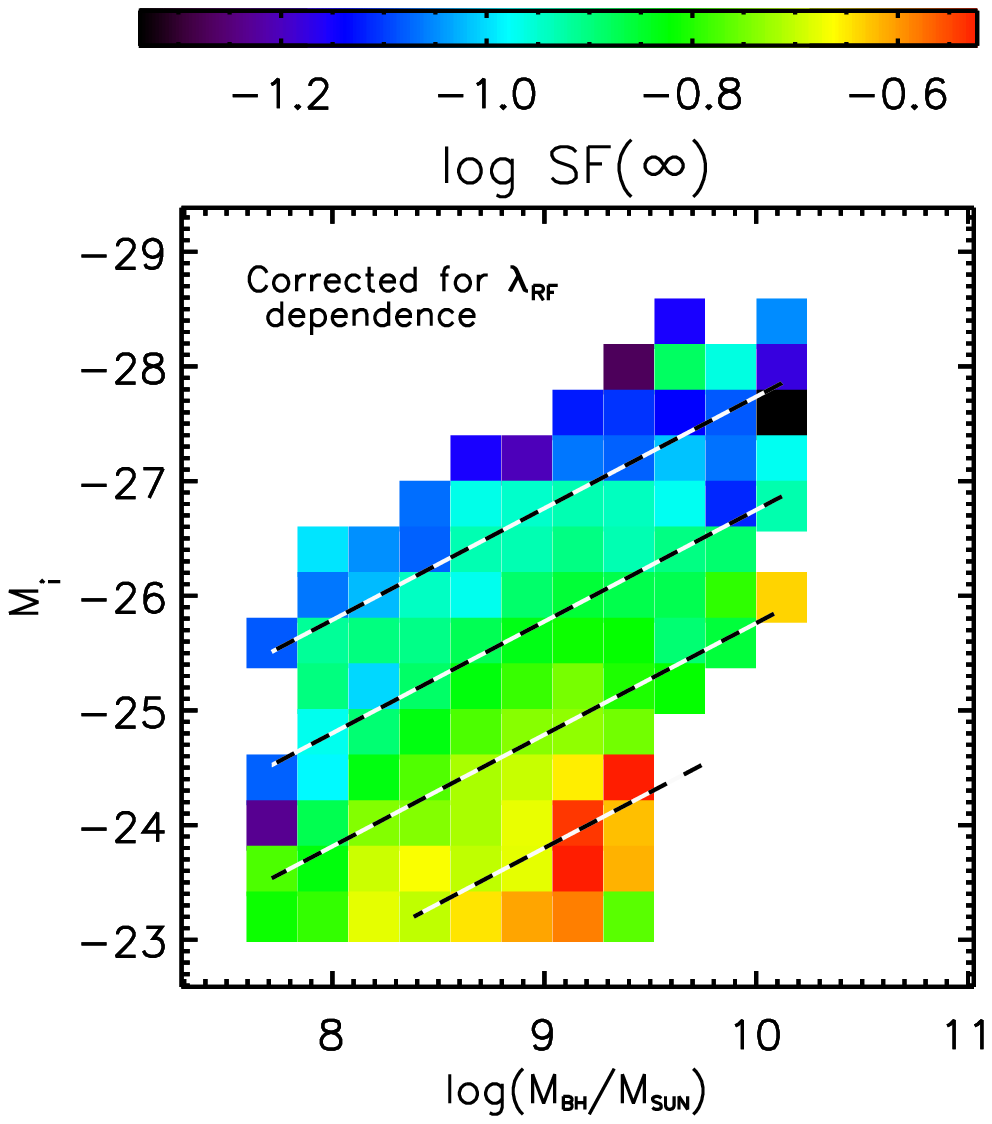}
     \includegraphics[width=2.5in]{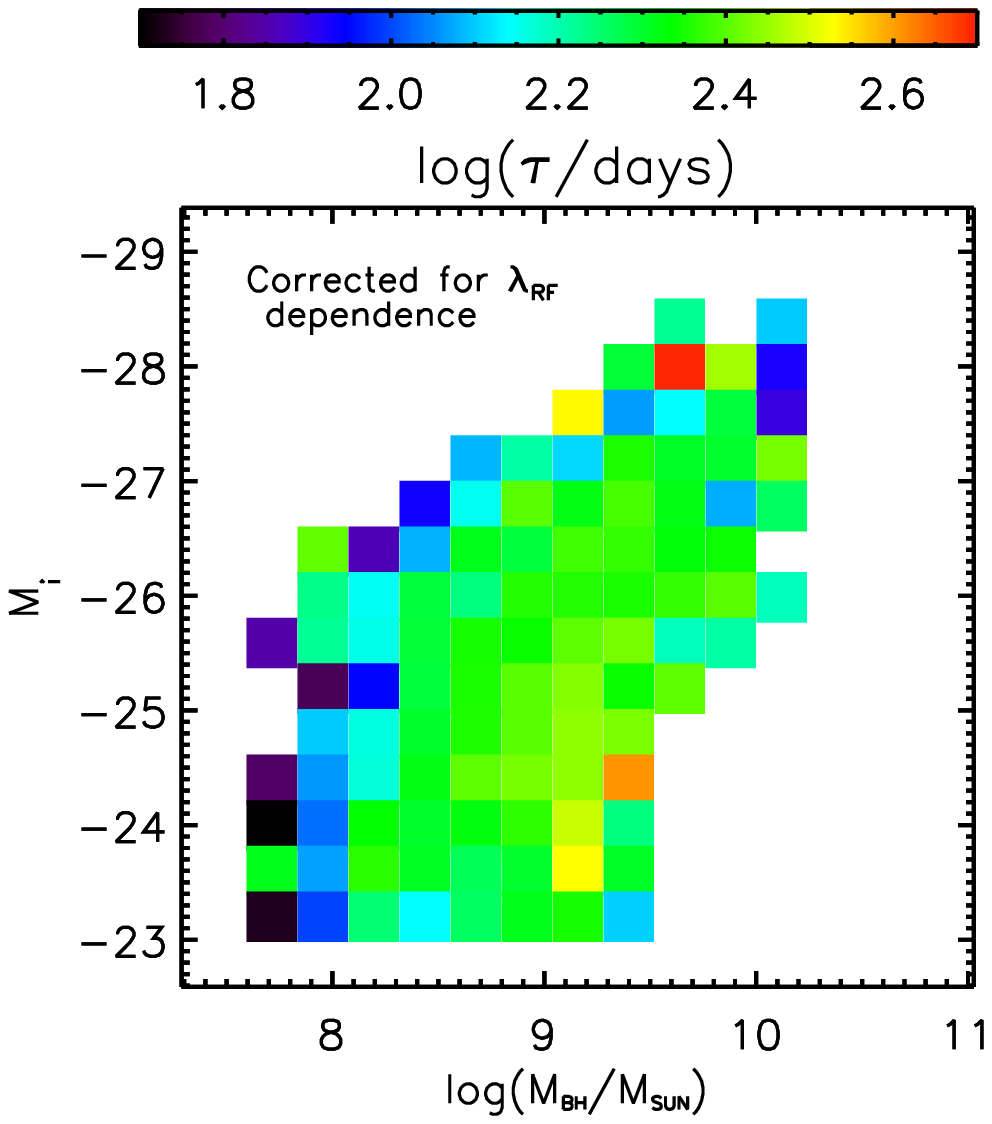}}
\caption{\emph{Top panels:} The long-term rms variability 
  $SF_{\infty}$ (\emph{left}) and characteristic time scale $\tau$
  (\emph{right}) are shown as colors on a grid of redshift and
  absolute i-band magnitude $M_i$. 
  The structure function parameters are normalized to a fixed rest
  wavelength using the fitted power law dependencies of
  $(\lambda_{RF}/4000\AA)^B$ with $B = -0.479$ and 0.17 for
  $SF_{\infty}$ and $\tau$, respectively.
  The lines of constant variability (dashed) show that $SF_{\infty}$
  is independent of redshift. \emph{Bottom panels:} As in the top
  panels but with black hole mass $M_{BH}$ on the x-axis.}
\label{fig:summary}
\end{figure*}

In the upper panels of 
Figure~\ref{fig:summary}, the median values of $SF_{\infty}$ and $\tau$
are shown as a function of absolute magnitude ($M_i$) and redshift. The structure function
parameters are normalized to a fixed rest wavelength of 4000\AA\ using the fitted
power law dependencies of $(\lambda_{RF}/4000\AA)^B$ with $B = -0.479$
and 0.17 for $SF_{\infty}$ and $\tau$, respectively, from Section~\ref{sec:trendslRF}, 
before finding the median in each pixel. For $SF_{\infty}$, 
the anti-correlation with luminosity clearly dominates any trend with
redshift. The $\tau$ distribution also shows negligible correlation with redshift.  
The bottom panels show the dependence on $M_{BH}$. 
Using a grid of $M_i$ vs. $M_{BH}$ allows one to search for trends in 
the variability parameters while accounting for the selection effect 
that more luminous galaxies host more massive black holes. 
A positive correlation between $SF_{\infty}$ 
and $M_{BH}$ is apparent, independent of the correlation with $M_i$. 
This is in agreement with the result from Wilhite et al.~(2008), who
used ensemble structure functions.
The $\tau$ parameter shows a clear correlation with $M_{BH}$, which
dominates any trend with $M_i$.

Motivated by these qualitative results, we fit a power law of the form 
\begin{eqnarray}
\log{f} = A + B\log\left(\frac{\lambda_{RF}}{4000{\rm \AA}}\right) + C(M_i+ 23) 
+ \nonumber \\ D\log\left(\frac{M_{BH}}{10^9M_{\odot}}\right) + E\log(1+z),
\label{eq:form}
\end{eqnarray}
to all $SF_{\infty}$ and $\tau$ data points in each $ugriz$ band separately, 
keeping $B$ fixed to $-0.479$ and $0.17$, respectively, in order to
avoid the bias discussed in Section~\ref{sec:trendslRF}. 
While there is a lot of scatter in the variability parameters for
fixed $M_i$, $M_{BH}$, and $\lambda_{RF}$ (see Section~\ref{sec:tests}), 
Eq.~\ref{eq:form} describes the mean trends well.  The best-fit
coefficients, averaged over the five bands, are reported in the first and
sixth rows of Table~\ref{tab:coeff}. The best-fit coefficients when
simultaneously fitting all bands are consistent with these
averages. The reported error bars are computed from the variation 
of the best-fit parameters over the five bands. 

The dependence of $\tau$ on redshift and $M_i$ is only marginally
detected, and can be attributed to the $L$--$z$ degeneracy: as
redshift increases, the best-fit $\tau$ decreases, while the
coefficient for $M_i$ indicates that as luminosity increases, $\tau$
increases. When $E$ is fixed to zero, the dependence on $M_i$ has low
significance, while the correlation with $M_{BH}$ remains the same
(seventh row of Table~\ref{tab:coeff}). For illustration, when $E$ is
fixed to 1 [so $\tau \propto (1+z)$, third and eighth rows], a
spurious dependence on $M_i$ emerges. 
Therefore, the controlling variable for the characteristic time scale
is clearly $M_{BH}$, suggesting that more massive black holes vary on
longer time scales. Similarly, with $E$ fixed to 0 for $SF_{\infty}$
(second row), the $C$ ($M_i$) and $D$ ($M_{BH}$) coefficients remain
largely unchanged, confirming that there is no significant correlation
between amplitude and redshift. Therefore, we force the final adopted
model to have no redshift dependence ($E=0$). Table~\ref{tab:coeff}
also provides fits for when $D$ is fixed to zero so that $SF_{\infty}$ and 
$\tau$ can be estimated in the absence of black hole mass information
(fourth and ninth rows).  

Of the analyzed parameters, the probable errors for black hole mass
estimates are the largest: of order 0.2 -- 0.4 dex (Marconi et al.~2008;
Vestergaard \& Peterson 2006).  When ignored, these large statistical
uncertainties result in underestimated values for $D$ (Kelly 2007). 
With an assumed random uncertainty of 0.2 dex in the black hole mass
measurements,  
we find the best-fit coefficients reported in the fifth and tenth
rows of Table~\ref{tab:coeff} (continuing to assume that $E=0$). 
It can be seen that including mass uncertainties of 0.2 dex increases
$D$ from $\sim$0.1 to $\sim$0.2 for both $\tau$ and $SF_{\infty}$ (a
bias of about 2-3 formal statistical fitting errors).  
{\it The coefficients in the fifth and tenth rows represent our
  best-fit model for the variability parameters, and we assume these
  values in the remaining Sections. } 

The best-fit model shows a smaller slope for the correlation between
$\tau$ and $M_{BH}$ of $0.21\pm 0.07$ than the value of $1.0\pm 0.4$
found by KBS09. The two results are still marginally statistically
consistent given the large uncertainties in KBS09. Moreover, their
sample contains relatively lower mass and luminosity quasars than
those analyzed in this study, and this difference might result in
additional biases. Koz10 also noted that the KBS09 results may be
affected by contamination from host galaxy emission, a problem which
will be far smaller for the generally more luminous SDSS quasars.

\begin{deluxetable}{c c c c c c}
\tabletypesize{\footnotesize}
\tablewidth{0pt}
\tablecaption{Best-Fit Coefficients for Eq.~\ref{eq:form} \label{tab:coeff}}
\tablehead{$f$ & $A$ & $B$ ($\lambda_{RF}$) & $C$ ($M_i$)& $D$ ($M_{BH}$)& $E$ ($1+z$)}
\startdata
$SF_{\infty}$                  &$-0.57 \pm 0.01 $&$-0.479\pm 0.005$&$ 0.117\pm 0.009$&$0.11\pm 0.02$&$0.07\pm 0.05$\\ 
$SF_{\infty}$                  &$-0.56 \pm 0.01 $&$-0.479\pm 0.005$&$ 0.111\pm 0.005$&$0.11\pm 0.02$&$\equiv 0$\\  
$SF_{\infty}$                  &$-0.760\pm 0.009$&$-0.479\pm 0.005$&$ 0.193\pm 0.006$&$0.12\pm 0.02$&$\equiv 1$\\  
$SF_{\infty}$                  &$-0.618\pm 0.007$&$-0.479\pm 0.005$&$ 0.090\pm 0.003$\tablenotemark{a}&$\equiv 0$&$\equiv 0$\\  
\hline 
$SF_{\infty}$\tablenotemark{c} &$-0.51\pm 0.02$&$-0.479\pm 0.005$&$ 0.131\pm 0.008$\tablenotemark{b}&$0.18\pm 0.03$&$\equiv 0$\\ 
\tableline 
$\tau$                  &$\phantom{-0} 2.4 \pm 0.2\phantom{0} $&$\phantom{-0} 0.17 \pm 0.02\phantom{0} $&$-0.05 \pm 0.03\phantom{0} $&$0.12\pm 0.04$&$ -0.7\pm 0.5\phantom{-}$\\
$\tau$                  &$\phantom{-0} 2.3 \pm 0.1\phantom{0} $&$\phantom{-0} 0.17 \pm 0.02\phantom{0} $&$\phantom{-} 0.01 \pm 0.03\phantom{0} $&$0.12\pm 0.04$&$\equiv 0$\\
$\tau$                  &$\phantom{-0} 2.1 \pm 0.1\phantom{0} $&$\phantom{-0} 0.17 \pm 0.02\phantom{0} $&$\phantom{-} 0.09 \pm 0.03\phantom{0} $&$0.13\pm 0.04$&$\equiv 1$\\
$\tau$                  &$\phantom{-0} 2.2 \pm 0.1\phantom{0} $&$\phantom{-0} 0.17 \pm 0.02\phantom{0} $&$-0.01 \pm 0.02\phantom{0} $&$\equiv 0$&$\equiv 0$\\
\hline 
$\tau$\tablenotemark{c} &$\phantom{-0} 2.4 \pm 0.2\phantom{0} $&$\phantom{-0} 0.17 \pm 0.02\phantom{0} $&$\phantom{-} 0.03 \pm 0.04\phantom{0} $&$0.21\pm 0.07$&$\equiv 0$\\
\tableline
\enddata
\tablenotetext{a}{Based on K-corrected magnitudes. Without the K-correction, this coefficent changes to $0.079\pm 0.003$.} 
\tablenotetext{b}{Based on K-corrected magnitudes. Without the K-correction, this coefficent changes to $0.113\pm 0.006$. 
}
\tablenotetext{c}{Measurement errors in $M_{BH}$ of 0.2 dex have been
  included in the fitting. These coefficients are recommended when applying the model. }
\tablecomments{In each row, the $B$ coefficient was determined and
  fixed before fitting a multiple regression in all other parameters  (see
  Section~\ref{sec:trendslRF}). The cosmology used for determining $M_i$
  is $\Omega_M = 0.30$, $\Omega_{\Lambda} = 0.70$, and $h=0.70$ (Schneider et al.~2007), 
  whereas that used in the $M_{BH}$ estimates is $\Omega_M = 0.26$,
  $\Omega_{\Lambda} = 0.74$, and $h=0.71$ (Shen et al.~2008). This  
  difference should only have a 1\% effect on the best-fit
  coefficients. }
\end{deluxetable}

\subsection{The Eddington Ratio as the Driver of Variability?}
\label{sec:Edd}
\begin{figure*}[h!]
\centerline{
\includegraphics[width=2.5in]{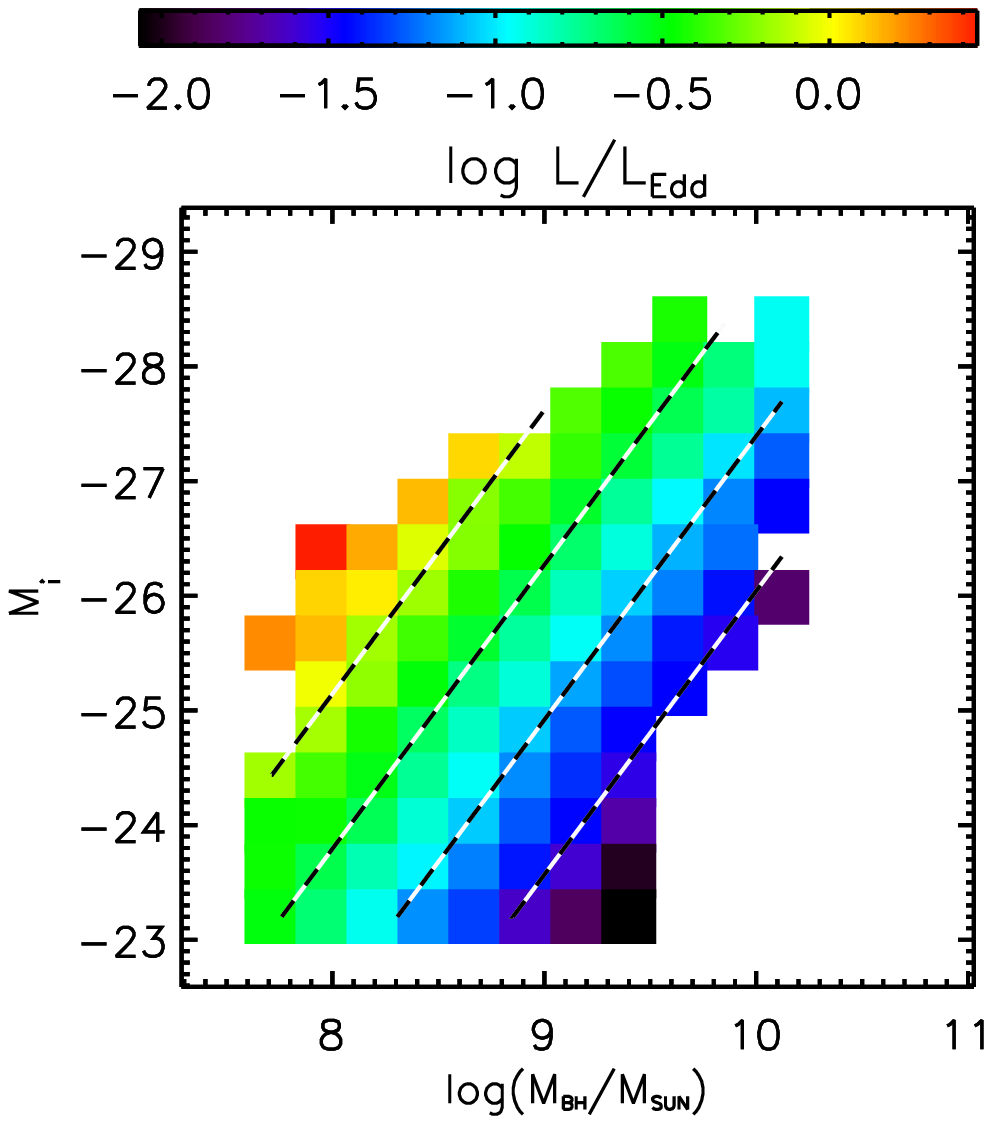}
\includegraphics[width=2.5in]{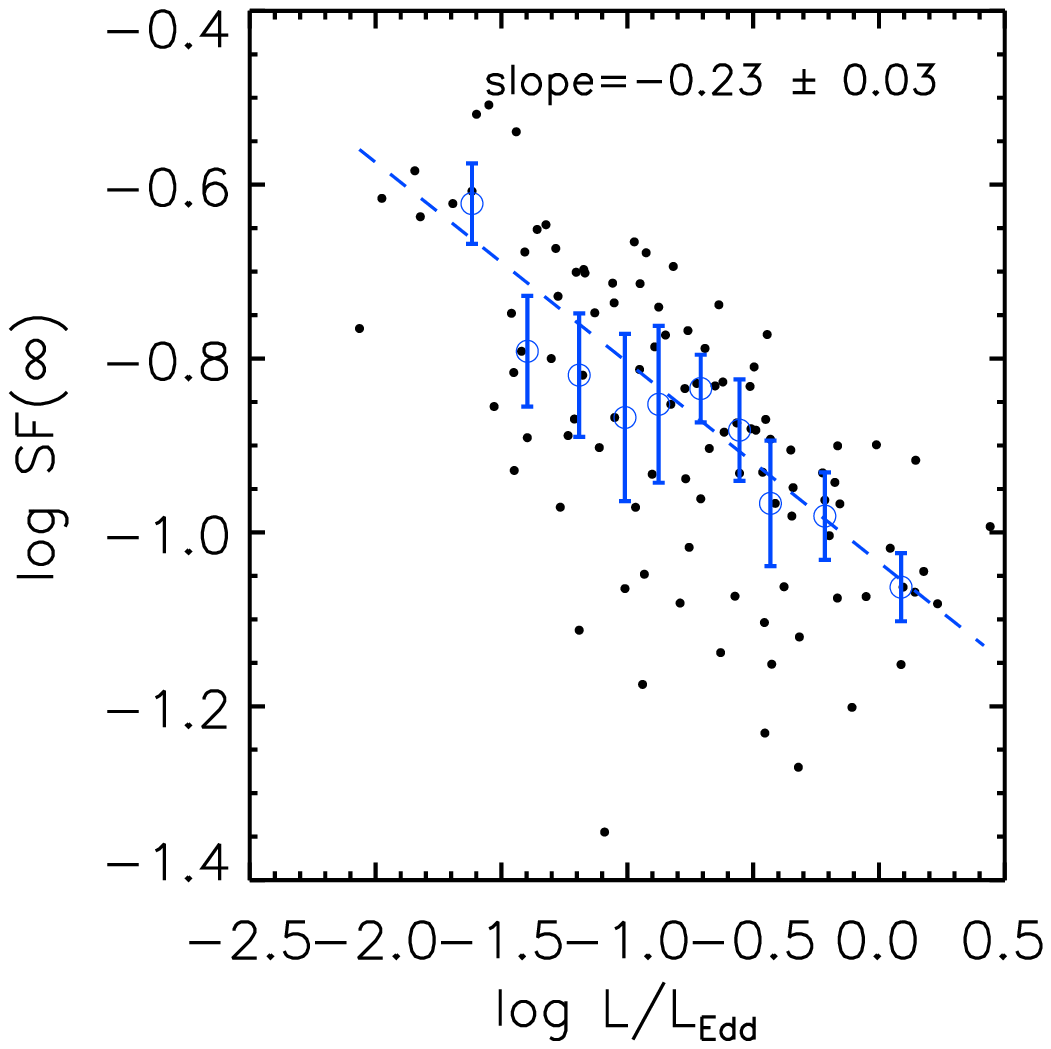}}
\caption{\emph{Left:} The Eddington ratio for S82 quasars (estimated
  using masses and bolometric luminosities from Shen et al.~2008) is shown as colors
  on a grid of $M_i$ vs.\ $M_{BH}$, with dashed lines of constant
  $L/L_{Edd}$ over-plotted. 
  \emph{Right:} Long-term rms variability (corrected for wavelength
  dependence) is shown as a function of
  $L/L_{Edd}$ (open circles are medians in each bin). The slope of the
  linear fit to the medians is listed on the panel.}
\label{fig:Edd}
\end{figure*}

Since the Eddington ratio, $L/L_{Edd}$, is dependent on luminosity and 
black hole mass, the trends for $SF_{\infty}$ in Table~\ref{tab:coeff}  
might be explained if $SF_{\infty}$ is simply driven by $L/L_{Edd}$.
To test this, we estimated $L/L_{Edd}$ as the ratio of the bolometric
luminosities from Shen et al.~(2008) to the Eddington luminosity,
$L_{Edd} = 1.5\times 10^{38}(M_{BH}/M_{\odot})$ erg/s.
Figure~\ref{fig:Edd} shows $L/L_{Edd}$ as a function of $M_{BH}$ and
$M_i$, and demonstrates that lines of constant $L/L_{Edd}$ are similar
in slope to those of constant $SF_{\infty}$ (see Figure~\ref{fig:summary}). 
The median $SF_{\infty}$ versus the median $L/L_{Edd}$ for each bin is
shown in the right panel, where we find a power law slope of $-0.23\pm
0.03$. This significant anti-correlation was also found by Wilhite
et al.~(2008), Bauer et al.~(2009), and Ai et al.~(2010). KBS09 did
not report such an anti-correlation; however, they did not compare
$L/L_{Edd}$ with $SF_{\infty}$ but rather with
$SF_{\infty}/\sqrt{\tau}$, the driving amplitude of short term
variability. If $L/L_{Edd}$ is the sole driver of the quasar
variability amplitude $SF_{\infty}$, we would expect that the
coefficients for $M_i$ and $M_{BH}$ are related as $D = 2.5C$. 
However, we find $D=(1.37\pm0.23)C$, 4.7$\sigma$ away from the
presumed slope of 2.5.  This suggests an additional source of
variability, such as a dependence on luminosity or on $M_{BH}$ in
addition to the dependence on $L/L_{Edd}$. Moreover, if we ignore this
additional source, the result that $SF_{\infty}$ depends on
$L/L_{Edd}$ supports a scenario where the amplitude of the optical
variability is determined by the accretion rate (see discussion in
Wilhite et al.~2008).  

If the Eddington ratio is a proxy for the quasar's age (e.g., Martini
\& Schneider 2003), then a lower Eddington ratio, and thus a larger
amplitude of variability, could indicate a dwindling fuel supply and a
more variable rate at which it is supplied to the black hole.
However, it is unlikely that the observed variability is due to
fluctuations in the external fuel supply to the disk, because the
fluctuation time scales of 10 -- 10,000 days are very short compared
to the viscous time scales of $10^5$ -- $10^7$ days that should
control large scale changes in the accretion rate. Moreover, these
fluctuations in the fueling rate will also be smoothed out and damped
as they travel inward, and will likely have effectively disappeared by
the time they reach the optical emitting region (e.g., see discussion
in Churazov et al.~2001). Instead, the origin of the fluctuations is
probably more local. 

Another possibility is that the dependence on $L/L_{Edd}$ may simply
be a reflection of the dependence on wavelength, which in turn depends
on the disk radius. If a higher $L/L_{Edd}$ means a hotter disk, then
the optical flux originates at a larger radius. Assuming that longer
wavelengths are emitted further out in the disk, the lower
$SF_{\infty}$ at longer wavelengths would lead to the anti-correlation
between $SF_{\infty}$ and $L/L_{Edd}$. In thin disk theory (e.g., see
Frank, King, \& Raine 2002), the characteristic radius for emission at 
wavelength $\lambda_{RF}$ scales as 
$R_{\lambda} \propto M_{BH}^{2/3} (L/L_{Edd})^{1/3} \lambda_{RF}^{4/3}$, 
and the thermal time scale is related to the orbital time scale as 
$t_{th} \propto t_{orb} \propto R^{3/2}/\sqrt{M_{BH}}$.
Therefore, under the assumption that $\tau$ is related to the thermal time 
scale, and that variability at wavelength  $\lambda_{RF}$ is 
dominated by the scale $R_{\lambda}$, 
$\tau$ scales with $\lambda_{RF}$, $M_{BH}$, and $L$ as: 
\begin{equation}
\tau \propto M_{BH}^{1/2} (L/L_{Edd})^{1/2} \lambda_{RF}^2.
\end{equation}
Since $L_{Edd} \propto M_{BH}$, this means $\tau$ scales as
\begin{equation}
\tau \propto L^{1/2} \lambda_{RF}^2,
\end{equation}
which does not match the observed scaling of 
$\tau \propto L^{-0.075} \lambda_{RF}^{0.17} M_{BH}^{0.21}$. 
If we assume that $\tau$ is related to the viscous time scale, 
\begin{equation}
\tau \propto L^{7/60} \lambda_{RF}^{5/3} M_{BH}^{2/3},
\end{equation}
still in conflict with the measured values. 

The variability time scale simply does not show the strong dependence 
on $\lambda_{RF}$, $M_{BH}$, and $L$ expected from these simple scalings. 
Therefore, using this naive scaling, we are not able to relate the
observed $\tau$ to either a thermal or viscous time scale of the
radius associated with the wavelength of the variability. However, in
reality there is a range of radii, which corresponds to a range in
time scales, contributing to the observed flux in each band, and this
will cause some degree of smoothing. Also, the radial regions might
overlap for each band, causing a single radius to contribute flux in
multiple bandpasses, and this might attenuate some of the dependencies
on $\lambda_{RF}$, $M_{BH}$, and $L$. Finally, our observed power law
indices may be biased due to the uncertainty in the bolometric
correction.

\subsection{A ``Recipe'' for Generating Mock Light Curves}
\label{sec:recipe}

Given the success of the damped random walk model in explaining a large 
body of observations, it represents a simple quantitative framework for
generating mock quasar light curves. The optical light curves can be
simulated for quasars at different redshifts and for a wide range of 
luminosity and black hole mass, which provides the basis for quantitative
modeling and optimization of quasar variability surveys. The model presented 
has already been implemented as a part of the simulation effort in support 
of the Large Synoptic Survey Telescope (Ivezi\'{c} et al. 2008). 

\begin{figure*}[h!]
\centering
\includegraphics[width=2.5in]{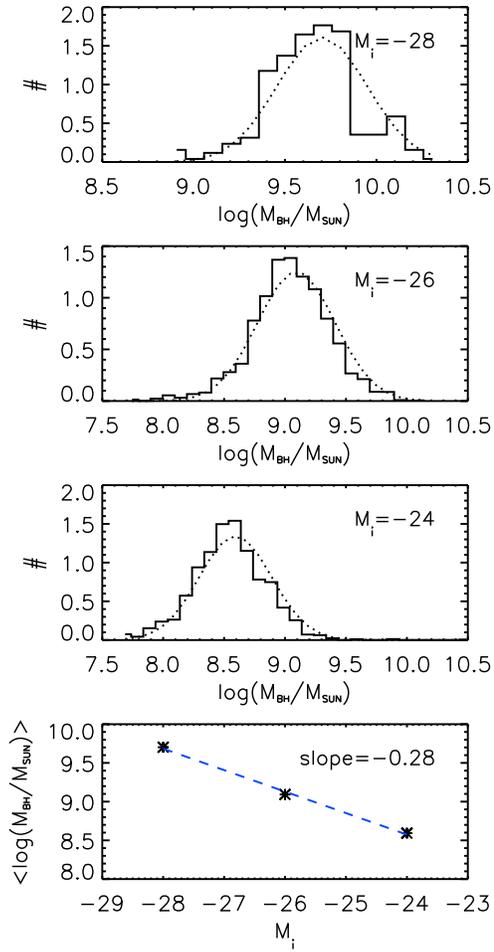}
\caption{Determination of $p(M_{BH} | M_i)$ (see
  Eq.~\ref{eq:pMBHMi}). The histograms in the top three panels show
  the distribution of black hole masses for S82 quasars with absolute
  magnitudes within $0.5$ mags of the value listed in the upper right
  corners. In the bottom panel, the mean value of
  $\log{(M_{BH}/M_{\odot})}$ from each 
  histogram is plotted against $M_i$. 
}
\label{fig:pxy}
\end{figure*}

At a given redshift, a simulated value for absolute magnitude can be readily drawn from 
an adopted luminosity function (e.g., Croom et al. 2009 and references
therein). An estimate for the black hole mass is also required in order to 
apply our model, and it can be generated using the adopted absolute magnitude
as follows. According to the Shen et al.~(2008) results, the quasar luminosity and black
hole mass are strongly correlated (see the bottom-right panel of
Figure~\ref{fig:hiddensys}). Using Shen et al.~(2008) values, we quantify this 
correlation in Figure~\ref{fig:pxy}. At a given $M_i$, the $M_{BH}$ 
distribution has a finite width due to the distribution of Eddington 
ratios and measurement errors. Both effects can be well-described using a 
Gaussian distribution:
\begin{eqnarray}
\label{eq:pMBHMi}
p(\log{M_{BH}} | M_i) = \nonumber \\
\frac{1}{\sqrt{2\pi}\sigma}\exp\left[-\frac{(\log{M_{BH}}-\mu)^2}{2\sigma^2}\right], 
\end{eqnarray}
where $\mu = 2.0 - 0.27M_i$ and $\sigma = 0.58 + 0.011M_i$ 
(the black hole mass is expressed in solar mass units). 

With adopted values for $M_i$ and $M_{BH}$, the variability amplitude 
and the characteristic time scale can be estimated using Eq.~\ref{eq:form}. 
We note that choosing a time scale from an adopted value for $SF_{\infty}$ 
by utilizing the correlation seen in Figure~\ref{fig:pardist} (and drawing 
from a Gaussian distribution similar to Eq.~\ref{eq:pMBHMi}) is not as accurate 
as using Eq.~\ref{eq:form} and the adopted values of $M_{BH}$ and $\lambda_{RF}$.
Finally, given a mean magnitude and wavelength, a quasar light curve can be
easily generated using Eq.~\ref{eq:car1sim}.

\section{Variability Properties of Radio- and X-ray--detected Quasars}
\label{sec:radio}

Previous studies, based mostly on two-epoch datasets, found that radio-loud 
quasars are marginally more variable than radio-quiet quasars for rest-frame 
time lags in the range 50--400 days (e.g., VB04 and references therein). 
It was also found that X-ray--detected quasars are significantly more variable 
than X-ray--undetected quasars at rest-frame time lags up to 250 days.

The large size and the availability of light curves for our sample allows 
us to revisit these results with more statistical power. In addition, the
damped random walk model and our best-fit results from Table~\ref{tab:coeff} 
provide a convenient method to account for various selection 
effects. For example, it is possible that subsamples selected by various means, 
such as the requirement of radio or X-ray detections, have different distributions 
of luminosity and black hole mass. In this case, they would display different 
variability behavior not because of intrinsic differences in the variability mechanism, 
but rather because of the trends captured by Eq.~\ref{eq:form}. A comparison of the 
{\it ratio} of observed and predicted variability parameters for any subsample and the 
full sample  will automatically take into account these sampling effects, and this is
the main statistical method we use in this section. We first discuss
a radio subsample, and then analyze a sample of quasars with X-ray detections.

\subsection{Radio-detected Subsample} 

\begin{figure*}[h!]
\centering
\includegraphics[width=2.5in]{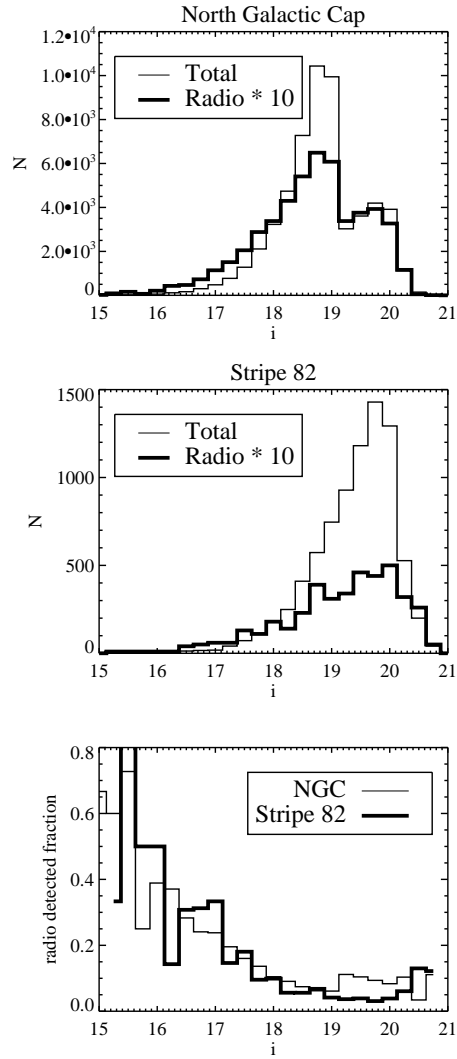}
\caption{\emph{Top:} The number of FIRST radio detections (thick line,
  multiplied by 10) as
  compared to the total number of quasars (thin line) as a function of $i$
  magnitude in the Northern galactic cap (NGC) footprint of the SDSS. 
  \emph{Middle:} As in the top panel but for Stripe 82.
  \emph{Bottom:} Fraction of quasars in the NGC (thin line) and in S82
  (thick line) samples that have radio detections. 
  Between $i=19$ and 20, the fraction of radio-detected quasars
  is considerably lower in S82 than in the NGC.}
\label{fig:radiodetect}
\end{figure*}

We use the unified radio catalog of Kimball \& Ivezi\'{c} (2008) to access the FIRST and 
NVSS data (20 cm continuum data)  for our objects. We refer the reader to this paper 
and references therein for details about these radio surveys and object association. 
Figure~\ref{fig:radiodetect} shows the radio-detected fraction of quasars in S82 as a 
function of magnitude. The overall fraction (5\%) is considerably lower than that in the 
Northern galactic cap footprint of the SDSS for the magnitude range $19<i<20$ 
(White et al.~1997; Ivezi\'{c} et al.~2002). This is due to a difference in targeting algorithms
between the two surveys, and results in nearly 300 quasars with both optical light curves and radio 
data. When available, we use NVSS values for radio flux, but otherwise 
we adopt the FIRST integrated flux. The radio-loudness parameter $R_g$ is calculated exactly 
as the $R_i$ in Ivezi\'{c} et al.~(2002), but using $g$ magnitudes instead. 
We assume a spectral index of $-0.5$ for the optical and 0 for the radio (because this
sample is dominated by the core radio emission; see Kimball \& Ivezi\'{c} 2008) when 
computing the K-correction.  

Table~\ref{tab:radio} compares the mean properties for various subsamples detected in the 
radio to the radio-undetected subsample (essentially the full sample due to small fraction
of radio-detected objects). For each subsample, the mean $\tau$ and
$SF_{\infty}$ are listed, as well as the mean ratio of the observed values to those predicted 
using Eq.~\ref{eq:form} and the measured values of $\lambda_{RF}$, $M_i$, and $M_{BH}$.
The mean values are computed iteratively with $\pm 3\sigma$ outliers excluded. 
Errors are reported as the (clipped) rms divided by $\sqrt{N-1}$, where N is the number of 
data points in the distribution.  Numbers of objects are lower for columns involving model 
quantities because black hole mass estimates are not available for all objects.

Table~\ref{tab:radio} shows that only the variability amplitude of the radio-loudest quasar 
subsample ($R_g \geq 3$) is significantly different ($>3\sigma$ deviation) from the behavior 
of the full sample. The radio-loudest quasars have systematically larger variability amplitudes,
when corrected for trends described  by Eq.~\ref{eq:form},  by about 30\%, compared to the
full sample dominated by radio-quiet objects, in agreement with VB04.

\subsection{X-ray Detected Subsample} 

For the analysis of variability properties of quasars detected at X-ray 
wavelengths, we use data from the ROSAT All-Sky Survey (RASS; Voges et al.~1999).
The X-ray subsample consists of 82 quasars with RASS full-band count rates greater 
than $10^{-3}$ ct/s, taken from the Schneider et al.~(2007) catalog.  As can be seen
from Table~\ref{tab:radio}, the variability properties of this subsample are statistically 
indistinguishable from the full sample. VB04 detected a significant increase in
structure function at rest time lags below 250 days for their X-ray subsample, 
but on long time scales there was no significant difference.  It is plausible 
that their result for short time scales was influenced by small sample size
and $\tau$ effects discussed in Section~\ref{sec:ringberg}.

\begin{deluxetable}{c c c c c c c c c}
\tabletypesize{\scriptsize}
\tablewidth{0pt}
\tablecaption{Mean Variability Properties of Radio and X-ray Subsamples \label{tab:radio}}
\tablehead{    & $\small{<}\log \tau\small{>}$&$N$&$\small{<}\log
  (\tau/\tau_p)\small{>}$&$N$&$\small{<}\log
  SF_{\infty}\small{>}$&$N$&$\small{<}\log
  (SF_{\infty}/SF_{\infty,p})\small{>}$&$N$}
\startdata
no radio       &$2.305\pm 0.005$&6989&$-0.003\pm 0.007$&4467&$-0.634\pm 0.003$&7067&$ 0.018\pm 0.003$&4508\\
no radio $i<19$&$2.327\pm 0.013$&1519&$-0.013\pm 0.013$&1279&$-0.721\pm 0.006$&1527&$ 0.003\pm 0.006$&1287\\
radio          &$2.26 \pm 0.03 $& 277&$-0.07 \pm 0.05 $& 154&$-0.70 \pm 0.02 $& 283&$-0.03 \pm 0.02\phantom{-} $& 154\\
radio $i<19$   &$2.24 \pm 0.06 $& 105&$-0.08 \pm 0.06 $&  73&$-0.81 \pm 0.03 $& 108&$-0.08 \pm 0.03\phantom{-} $&  74\\
$R_g \geq 3$   &$2.30 \pm 0.05 $& 135&$-0.01 \pm 0.07 $&  68&$-0.57 \pm 0.02 $& 137& ${\bf  0.11 \pm 0.03} $ & {\bf  68}\\
$R_g < 3$      &$2.28 \pm 0.03 $& 211&$\phantom{-} 0.00 \pm 0.04 $& 133&$-0.73 \pm 0.02 $& 219&$-0.06 \pm 0.02\phantom{-} $& 140\\
$R_g < 2$      &$1.99 \pm 0.14 $&  30&$-0.21 \pm 0.14 $&  19&$-0.88 \pm 0.07 $&  30&$-0.12 \pm 0.06\phantom{-} $&  19\\
resolved\tablenotemark{a}       &$2.29 \pm 0.05 $& 111&$\phantom{-} 0.00 \pm 0.06 $&  63&$-0.63 \pm 0.02 $& 111&$ 0.04 \pm 0.03 $&  63\\
unresolved\tablenotemark{a}     &$2.24 \pm 0.04 $& 167&$-0.13 \pm 0.07 $&  91&$-0.73 \pm 0.03 $& 172&$-0.06 \pm 0.03\phantom{-} $&  91\\
x-ray          &$2.41 \pm 0.05 $&  81&$\phantom{-} 0.03 \pm 0.06 $&  59&$-0.60 \pm 0.03 $&  82&$ 0.01 \pm 0.03 $&  58\\
no x-ray       &$2.307\pm 0.005$&6950&$-0.004\pm 0.007$&4559&$-0.638\pm 0.003$&7020&$ 0.017\pm 0.003$&4598\\
\tableline
\enddata
\footnotesize                                                          
\tablenotetext{a}{The morphological radio classes are defined using the
integrated and peak FIRST fluxes as in Kimball \& Ivezi\'{c} (2008).}
\tablecomments{$\tau$ and $SF_{\infty}$ are the observed time
  scales and asymptotic amplitudes of optical variability
  (Section~\ref{sec:results}), while $\tau_p$ and $SF_{\infty,p}$
  refer to those predicted from Eq.~\ref{eq:form}, using the
  coefficients in the fifth and the tenth rows of Table~\ref{tab:coeff}.
  Errors are reported as the rms divided by $\sqrt{N-1}$, where
  N is the number of data points, listed after each column. Numbers of
  objects are lower for columns involving model estimates because 
  black hole mass estimates are not available for all objects. Only the
  variability amplitude for  the radio-loudest quasar subsample ($R_g \geq 3$)
is significantly different ($>3\sigma$ deviation) from the behavior of 
the full sample.}
\end{deluxetable}

\section{Summary and Conclusions}
\label{sec:discussion}
We have used the damped random walk model of KBS09 and Koz10 to model
the optical/UV variability of $\sim$9000 SDSS Stripe 82 quasars with the
$ugriz$ light curves. The dataset includes 2.7 million photometric measurements 
collected over 10 years.  We confirm that this is a good model of quasar
    variability, and quantify the dependence of two variability
    parameters, the long-term rms variability $SF_{\infty}$
and the damping time scale $\tau$, on physical parameters such as wavelength, luminosity,
   black hole mass and Eddington ratio.  
Our main results are the following:
\begin{enumerate}
\item A stochastic process with an exponential covariance function
  characterized by an amplitude and time scale provides a good
  fit to observed quasar light curves, as shown by KBS09 and Koz10, using smaller 
  samples with less wavelength coverage, but better time sampling. 
\item The long-term rms variability, $SF_{\infty}$, has a mode at $\sim$0.2 mag and
  characteristic time scales, $\tau$, are roughly 200 days in the rest
  frame, as 
  found previously by KBS09 and Koz10. These time scales are consistent
  with thermal time scales, but simple accretion disk models fail to reproduce
  the observed scaling of $\tau$ with physical parameters. 
\item
  Quasars with similar physical parameters can have different 
  characteristic time scales for variability. 
  It is now clear that the
  distribution of $\tau$ accounts for most of the scatter in the structure function on short
  time scales for quasars with similar 
  luminosity, rest wavelength, and time lag, which explains the puzzling results from MacLeod et al.~(2008).  
  Results from fitting a power law to observed ensemble structure functions should be
  interpreted with caution. 
\item The variability time scale is correlated with the long-term rms 
  variability with a slope of $1.30\pm0.01$ dex/dex. Quasars that have
  large long-term variability amplitudes generally vary on longer characteristic time scales. 
  The amplitude of short term variations is also correlated with $\tau$.
  This conclusion is unaffected by any time sampling issues in the S82 dataset.
\item The damped random walk corresponds to a PSD proportional to
  $1/f^2$ at frequencies $f>(2\pi\tau)^{-1}$, flattening to a constant
  at lower frequencies. At large $f$, the data
  are in great agreement with $PSD \propto 1/f^2$. In terms of the
  structure function, this means that $SF \propto (\Delta
  t)^{1/2}$. Whereas previous analyses of the SF obtained a
  power law slope of $\beta = 0.3$, here we demonstrated that this
  would be a consequence of fitting the data around the ``knee'' 
  (turn-over) of the SF. Our constraints for small $f$ are much
  weaker. As discussed in Section~\ref{sec:PSD}, due to a lack of
  sufficient long-time scale information, we are unable to distinguish  
  between a $1/f^{0}$ or a $1/f$ PSD at frequencies $f<(2\pi\tau)^{-1}$
  using the data and computational technique described here.  
\item The rest-frame variability parameters show a negligible trend with
  redshift, suggesting that they are intrinsic to the quasars,
  and these properties do not evolve over cosmic time
  for fixed physical parameters of the quasar ($M_{BH}$, $M_i$, and $\lambda_{RF}$).
\item For fixed luminosity and black hole mass, $\tau$ increases with
  increasing rest-frame wavelength with a power law
  index of $0.17$, and $SF_{\infty}$ decreases with a power law
  index of $-0.48$. The latter result is similar to previous findings (e.g., MacLeod et
  al.~2008, VB04). Koz10 also observed that the variability increases to shorter 
  $\lambda$, but they kept $\tau$ fixed in their fits. If wavelength is a proxy for radius in the
  accretion disk, this implies that the characteristic time scales are longer and the variability 
  amplitudes are smaller in the outer regions than in the inner regions. 
\item The long-term variability $SF_{\infty}$ is strongly
  anti-correlated with luminosity as found in previous studies such as  
  VB04, Wilhite et al.~(2008), and references therein. By studying the 
  median $SF_{\infty}$ in the plane of absolute magnitude and black hole 
  mass, we can separate the anti-correlation of amplitude with luminosity from 
  the positive correlation with black hole mass. As suggested in
  Wilhite et al.~(2008), these trends may be largely explained if the amplitude of
  variability is tied to changes in the accretion rate in the disk, 
  and is simply related to the Eddington ratio. However, despite the strong anti-correlation 
  between $SF_{\infty}$ and $L/L_{Edd}$ (which accounts for
  most of the dependence on $M_i$ and $M_{BH}$), the 
  exact dependence with $M_i$ and $M_{BH}$ is not consistent with 
  $L/L_{Edd}$ as the sole driver of quasar variability.
\item The damping time scale $\tau$ appears to be nearly independent of
  luminosity and correlated with $M_{BH}$ with a power law index
  of $0.21\pm 0.07$.  The mild discrepancy with the KBS09 result (1.0$\pm$0.4) may be due to 
  the different range of sampled luminosity and black hole mass, as well as 
  contamination by host galaxy emission in many of the very low
  luminosity systems they consider (see Koz10). 
\item While the mean variability parameters can be related to physical parameters, for 
  fixed values of $M_i$, $\lambda_{RF}$, and $M_{BH}$,  there is still a large
  scatter around the mean values, similar to the variance of the observed distributions.
  Some of that scatter can be attributed to measurement and fitting errors 
  ($\sim$60\% for $\tau$  and $\sim$70\% for $SF_{\infty}$), but there is
  enough evidence for residual stochastic nature of quasar variability. 
  Therefore, it cannot be assumed that quasars with similar $M_i$, $\lambda_{RF}$, and $M_{BH}$ 
  will necessarily have similar variability properties. 
\item 
The radio-loudest quasars have systematically larger variability amplitude by about 30\%, while 
the distribution of their characteristic time scale is indistinguishable from that of the full sample. 
There are no statistically robust differences in the characteristic time scale and variability
amplitude between the full sample and a small subsample of quasars detected by ROSAT.  
\end{enumerate}

With this paper, and results from KBS09 and Koz10, the ability
of the damped random walk model to quantitatively describe quasar 
variability is well-established.  As emphasized by Koz10,
this means that variability studies can become fully quantitative because
the entire process of identifying and assigning parameters to quasars
can be simulated to allow estimates of completeness and parameter
biases.  In particular, an important next step is to determine the
variability equivalents of luminosity functions, i.e., the intrinsic
distributions of the variability parameters.  While our results 
represent a good first step in this direction, we caution that 
a non-negligible fraction of the S82 quasars have indeterminately 
long time scales.  If, following KBS09, we identify the characteristic time
scale with the thermal time scale, then the next question is whether
there are additional time scales (such as the dynamical or viscous time scale),
or sources of variability. 
For example, Blackburne \& Kochanek~(2010) recently found evidence for
changes in disk size with changes in luminosity using gravitational
microlensing. Such searches for additional sources of variability will likely 
require better sampled light curves and over a longer time scale. 

The prospect of advancing these studies of quasar variability now faces a bottleneck. 
The S82 quasars have the advantage of sample size, wavelength coverage, 
and spectroscopy, but the light curves have poor sampling and modest
overall lengths, leading to significant problems for accurately 
estimating $\tau$ when it is long.  The quasars behind the Magellanic 
Clouds (Koz{\l}owski \& Kochanek 2009) are a smaller sample without spectroscopic confirmation, but
have superb, long term light curves that continue to be extended because 
of the continuing microlensing projects.  Improving on our present
results in the short term depends on either reviving the monitoring
of S82 or spectroscopically confirming the Magellanic Cloud quasars.

Resuming the monitoring of Stripe 82 is challenging with the 
decommissioning of the SDSS imaging system.  The best short-term 
prospects are the Pan-STARRS project (Kaiser et al.~2002), the Palomar-QUEST 
project  (Schweitzer et al.~2006), or using the DECam being built for the Dark 
Energy Survey (Honscheid et al.~2008).  Since the challenge
is to constrain long time scales, the presence of a multiple-year gap
is mainly a complication for ensuring that problems in matching 
photometric bands are not interpreted as a form of long-term 
variability.  
Obtaining spectra of the Magellanic Cloud quasars is in 
some ways easier, because the quasar magnitudes and densities are
well-suited to the AAOmega fiber spectrograph on the AAT and, to
a lesser degree, the IMACS spectrograph on Magellan.

\begin{figure*}[p]
\centering
\includegraphics[width=5in]{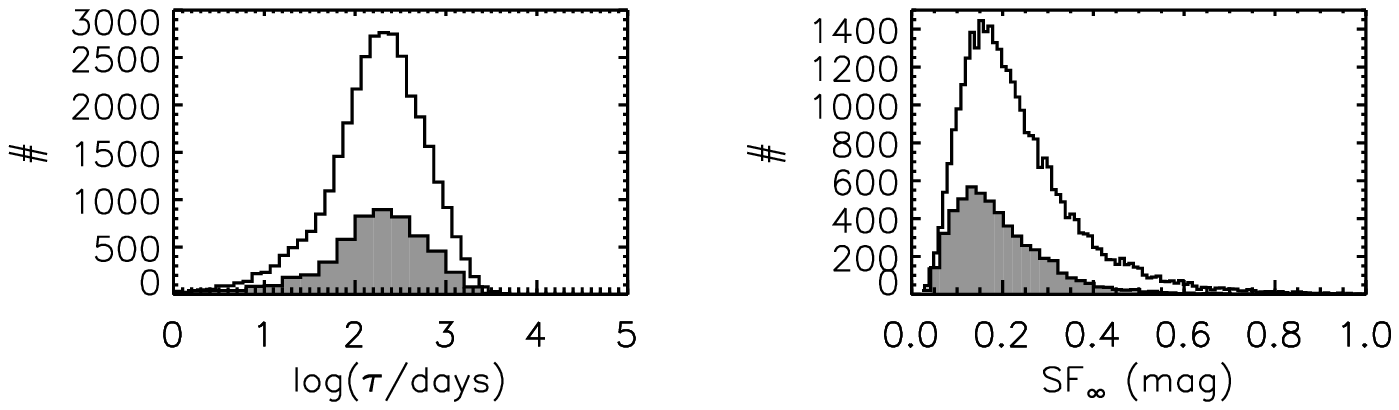}
\includegraphics[width=5in]{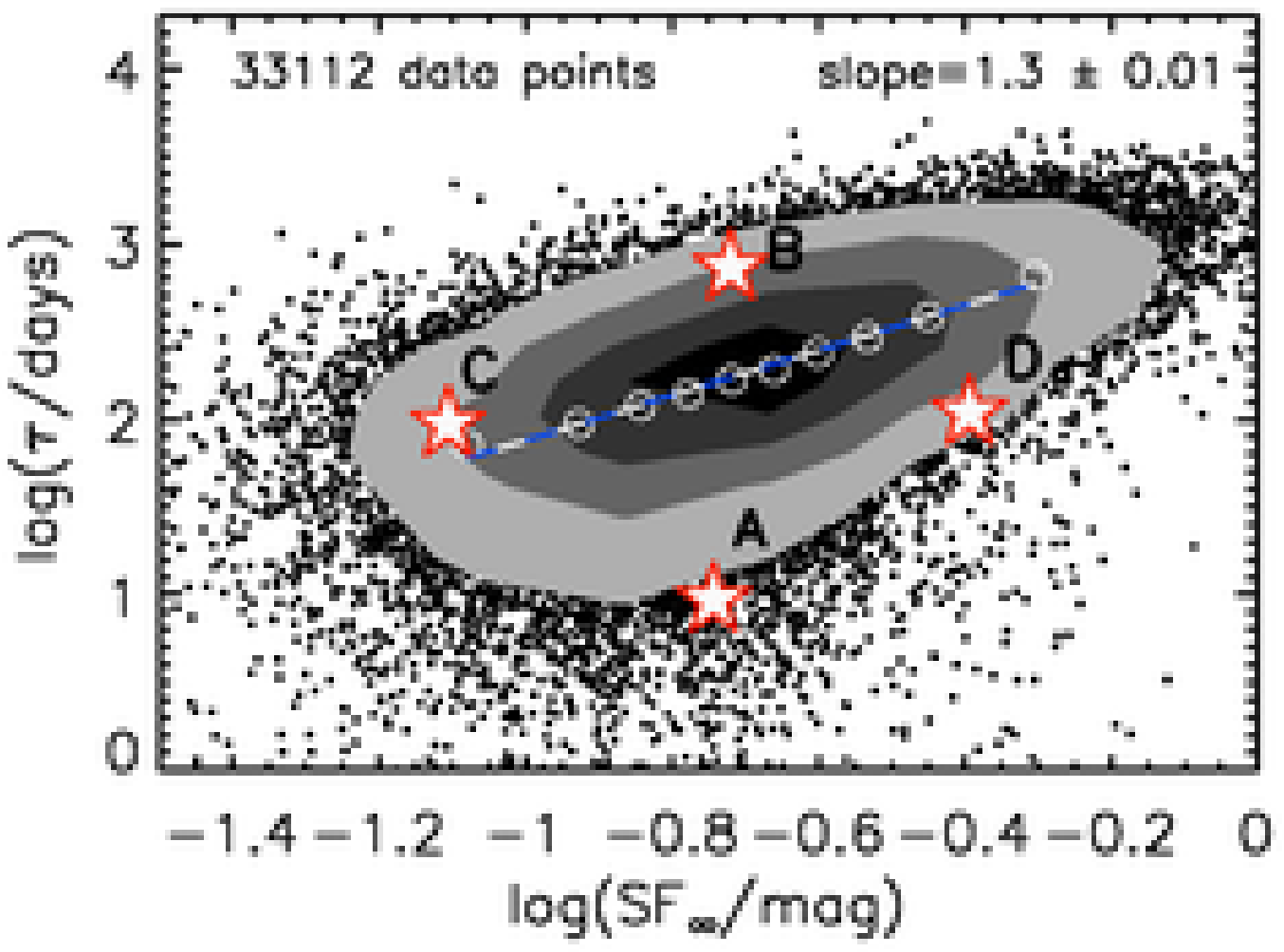}
\caption{\emph{Top}: Distributions of the best-fit 
   rest-frame time scale $\tau$ (\emph{left})
   and long-term structure function $SF_{\infty}$ (\emph{right}) for the
   S82 quasars ($33,112$ light curves). 
   The filled histograms show the fits for quasars with absolute
   i-band magnitude in the range $-27<M_i<-26$ (the median $M_i$ for the
   whole sample is $-25.46$). 
\emph{Bottom}: Relationship between $\tau$ and
  $SF_{\infty}$. A power law shown by the dashed line is fit to the medians (open
   white circles), with the slope listed in the top-right
   corner. Contours show regions containing 90\%, 70\%, 50\%,
   and 20\% of the total number of points. The star symbols label the
   four regions in parameter space from which the sample light curves in
   Fig.~\ref{fig:LCeg1} (labeled A through D) were chosen. The observed-frame $\tau$ distribution
   lacks many of the short time scales observed in Koz10. This is
   likely due to either the better time sampling of the OGLE light
   curves or stellar contamination in their sample.}
\label{fig:pardist}
\label{fig:pr_tau_sf}
\end{figure*}

\begin{figure*}[p]
\centering
\includegraphics[width=7in]{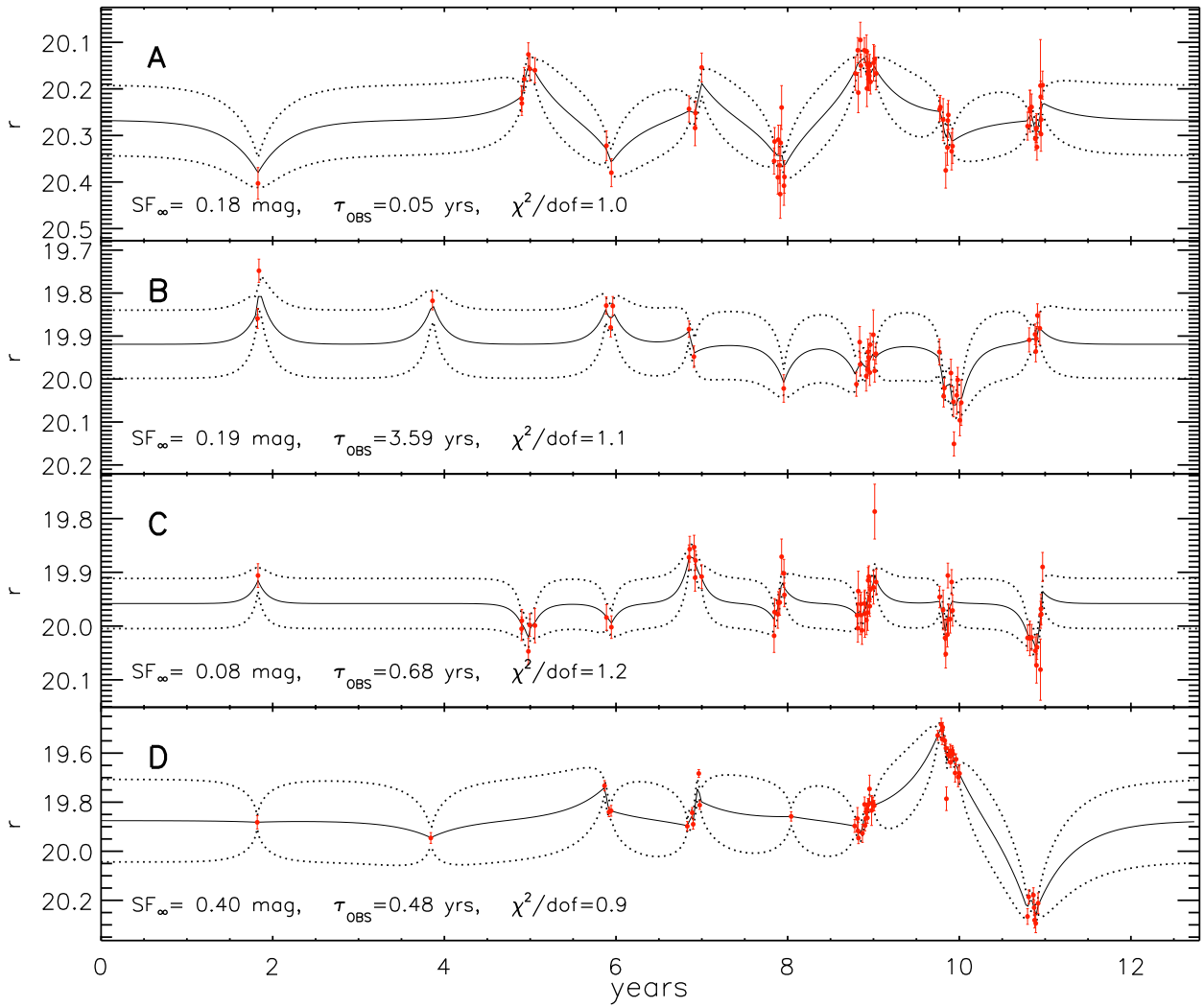}
\caption{Each panel shows a sample light curve from a certain region
  of $\tau$, $SF_{\infty}$ space (the four regions A--D are indicated with
  stars in the bottom panel of
  Figure~\ref{fig:pr_tau_sf}). Data points with error bars show the
  observed S82 data. Solid lines show the weighted average of the damped random
  walk model light curves that are consistent with the data (see Section~\ref{sec:methodology}). Dotted
  lines show the $\pm 1\sigma$ range of these possible stochastic
  models about the average.}
\label{fig:LCeg1}
\end{figure*}

The more general, Northern monitoring projects such as Pan-STARRS 
and Palomar-QUEST will slowly build light curves for essentially all the 
SDSS quasars, but at present their cadences are not ideal (see Schmidt et al.~2010) and it 
will take nearly a decade to build the long duration light curves needed 
for the analysis. 
In the long term, observations will 
be significantly improved with the advent of next-generation sky surveys. 
Most notably, the Large Synoptic Survey Telescope  (LSST, Ivezi\'{c} et al.
2008) will obtain accurate, well-sampled light curves for millions of
AGN. The observed distribution of rest-frame characteristic time scales for
S82 quasars spans the range from about 10 days to 1000 days (c.f.\
Figure~\ref{fig:pardist}).  To probe the time scales as short as
$0.1\tau$, and assuming a characteristic redshift of 2, the light
curves should be sampled every 3 days in the observer's frame, which
is in good agreement with the baseline cadence of LSST. With a 10
year-long survey, the length of the light curves will be in the range
(1-200)$\tau$.  
A combination of the SDSS, Pan-STARRS, DES, and LSST data for
$\sim$10,000 Stripe 82 quasars would span well over two decades, with
multi-band photometry obtained for hundreds of epochs, and would
represent the best sample to date for studying the optical continuum variability 
of quasars. In particular, such a dataset would enable a robust measurement
of the low-frequency behavior of their PSD (c.f.\ point 5 above).
For illustration, the LSST photometric errors in the $r$ band will be $<0.02$
mag for $r<22$, and there are roughly 2-3 million AGN with $r<22$ in
the 20,000 sq.\ deg.\ covered by the main LSST survey (see Table
10.2 in the LSST Science Book; Abell et al.~2009).  
Each of these objects will be observed about 1000 times,
yielding a database of over 2 billion photometric 
measurements. This data set, roughly a thousand times larger than that analyzed
here, will enable a significant improvement in our understanding of
quasar variability.

\acknowledgments

We acknowledge support by NSF grant AST-0807500 to the University of
Washington, and NSF grant AST-0551161 to LSST for design and development activity.
CSK and SK acknowledge support by NSF grant AST-0708082.
BK acknowledges support by NASA through Hubble
  Fellowship grant \#HF-51243.01 awarded by the
  Space Telescope Science Institute, which is operated by the
  Association of Universities for Research in Astronomy, Inc., for
  NASA, under contract NAS 5-26555. We thank an anonymous referee for
  valuable suggestions regarding the analysis in Section~\ref{sec:PSD}.

    Funding for the SDSS and SDSS-II has been provided by the Alfred P.\ Sloan Foundation, the Participating Institutions, the National Science Foundation, the U.S.\ Department of Energy, the National Aeronautics and Space Administration, the Japanese Monbukagakusho, the Max Planck Society, and the Higher Education Funding Council for England. The SDSS Web Site is http://www.sdss.org/.

    The SDSS is managed by the Astrophysical Research Consortium for the Participating Institutions. The Participating Institutions are the American Museum of Natural History, Astrophysical Institute Potsdam, University of Basel, University of Cambridge, Case Western Reserve University, University of Chicago, Drexel University, Fermilab, the Institute for Advanced Study, the Japan Participation Group, Johns Hopkins University, the Joint Institute for Nuclear Astrophysics, the Kavli Institute for Particle Astrophysics and Cosmology, the Korean Scientist Group, the Chinese Academy of Sciences (LAMOST), Los Alamos National Laboratory, the Max-Planck-Institute for Astronomy (MPIA), the Max-Planck-Institute for Astrophysics (MPA), New Mexico State University, Ohio State University, University of Pittsburgh, University of Portsmouth, Princeton University, the United States Naval Observatory, and the University of Washington.

{\it Facilities:} \facility{SDSS}, \facility{FIRST}, \facility{NVSS}, \facility{RASS}.

\clearpage

\appendix{{\bf Appendix: Search for Periodic Light Curves}}

Although a stochastic process has proven to
be an accurate statistical description of quasar light curves,
any discovery of periodic behavior (even of a single source) would have interesting physical
implications. 
However, a periodogram can only be used as an indicator of significant periodicity   
 in a signal as compared to pure white noise (i.e., having no signal at
all). Since quasars are genuinely variable, as described by a damped random walk 
(i.e., a red noise process; see KBS09), 
one can only evaluate the significance of the periodicity when also allowing for the signal 
covariance that is also present (see Markwardt et al.~2009; Gotz et al.~2009; Cenko et al.~2009, 
and references therein). 
Nevertheless, as a quick test for outstanding cases of periodicity among the
S82 light curves, we analyzed 8,863 light curves for evidence of periodicity using 
the Lomb-Scargle periodogram (Lomb 1976; Scargle 1982). 
The threshold
for considering the strongest peak in the periodogram as candidate evidence
for periodicity was set following Horne \& Baliunas (1986), with an
adopted false alarm probability of 0.05. However, this is the
threshold for ruling out white noise in favor of periodicity, and
therefore provides no information on how a periodic description compares to a
colored noise process such as a damped random walk. To determine
the latter, a different threshold is needed (see Koz10). When adopting 
the threshold for ruling out white noise in favor of periodic variability, 
we identify 88 light curves as good candidates. 

\begin{figure*}[h!]
\centering
\includegraphics[width=6in]{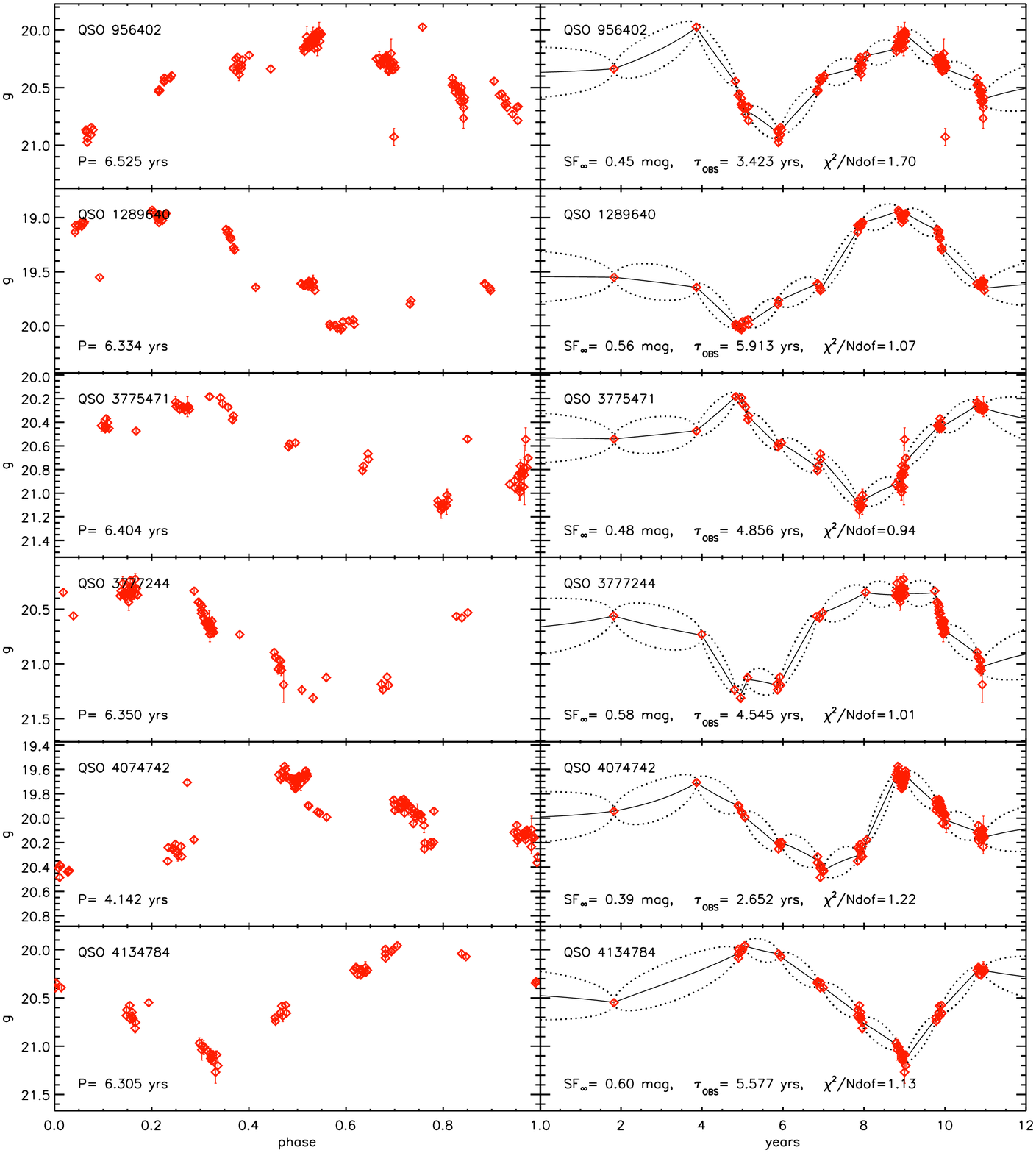}
\caption{\footnotesize Six examples of candidate light curves showing significant
  periodicity as compared to pure noise (see Appendix).  The phased
  light curves are shown on the \emph{left}, where the 
  phase is the fractional part of the ratio of time to the period (i.e.,
  the phase zero-point is arbitrary). The best-fit period is listed on
  the bottom of each panel. The full light curves are shown
  on the \emph{right}, where the solid lines show the weighted average
  of all consistent damped random walk model light curves, and the
  dotted lines show the $\pm 1\sigma$ range of possible stochastic 
  models (see Section~\ref{sec:methodology}). The best-fit damped
  random walk parameters are listed at the bottom of the right panels,
  along with the reduced $\chi^2$.}
\label{fig:periodic_phg}
\end{figure*}

A close inspection of the period distribution for the full sample and the 
selected 88 candidates shows important differences: while the full
sample displays a fairly flat distribution ranging from 100 days to values
exceeding 10,000 days, the period distribution for 88 candidates is 
bimodal. The first peak with 22 objects corresponds to aliasing at 
roughly one-year sampling cadence, while the second peak is centered
on periods of about 6-7 years, similar to the total length of observations. 
We have visually inspected the light curves and phased light curves for all
88 candidates. It turns out that candidates with proposed periods of
the order one year have light curves consistent with aliasing, while those
with longer periods typically have only one observed ``oscillation'' 
that might not be used as robust evidence for periodicity (six 
examples are shown in Figure~\ref{fig:periodic_phg}). 
Therefore, our
search for  periodic light curves in the S82 quasar sample has not
yielded any convincing cases.  
Again, it is important to note that in cases such as in Figure~\ref{fig:periodic_phg}, 
the periodogram indicates that these sources with long $\tau$ 
are likely to be periodic, not because they show true periodicity, but 
because it computes the likelihood relative to the wrong null hypothesis 
(white noise rather than colored noise). 
Moreover, the fact that $<$5\% of the light curves exceed the threshold
for periodicity being a better description than pure white noise is a
further indication that the periodogram is not a very powerful
statistic for poorly sampled light curves, since it is clear from the
previous Sections that quasar light curves are not white noise, but
rather are well described by a damped random walk. 
In fact, the single ``oscillation'' observed for the best candidate
long period objects is entirely expected from colored noise processes
such as a damped random walk (see right panels of
Figure~\ref{fig:periodic_phg}). Therefore, this analysis is further
evidence that a damped random walk is a good description of
quasar light curves.

\newpage

\begin{thebibliography}{}
\bibitem[Abazajian et al.(2009)]{2009ApJS..182..543A} Abazajian, K.~N., et 
al.\ 2009, \apjs, 182, 543 
\bibitem[LSST Science Collaborations: Paul A.~Abell et 
al.(2009)]{2009arXiv0912.0201L} Abell, P.~A.\ et al.\ 2009, arXiv:0912.0201;
  http://www.lsst.org/lsst/scibook 
\bibitem[Ai et al.(2010)]{2010arXiv1005.0901A} Ai, Y.~L., Yuan, W., Zhou, 
H.~Y., Wang, T.~G., Dong, X.~-., Wang, J.~G., 
\& Lu, H.~L.\ 2010, arXiv:1005.0901 
\bibitem[Aretxaga et al.(1997)]{are97} Aretxaga, I., Cid
Fernandes, R., \& Terlevich, R.~J.\ 1997, \mnras, 286, 271
\bibitem[Ar{\'e}valo et al.(2008b)]{2008MNRAS.389.1479A} Ar{\'e}valo, P., 
Uttley, P., Kaspi, S., Breedt, E., Lira, P., 
\& McHardy, I.~M.\ 2008b, \mnras, 389, 1479 
\bibitem[Ar{\'e}valo et al.(2008a)]{2008MNRAS.388..211A} Ar{\'e}valo, P., 
McHardy, I.~M., \& Summons, D.~P.\ 2008a, \mnras, 388, 211 
\bibitem[Bauer et al.(2009)]{bau09} Bauer, A., Baltay, C.,
Coppi, P., Ellman, N., Jerke, J., Rabinowitz, D.,
\& Scalzo, R.\ 2009, \apj, 696, 1241
\bibitem[Bhatti et al.(2010)]{2010ApJS..186..233B} Bhatti, W.~A., Richmond, 
M.~W., Ford, H.~C., \& Petro, L.~D.\ 2010, \apjs, 186, 233 
\bibitem[Blackburne 
\& Kochanek(2010)]{2010arXiv1002.3126B} Blackburne, J.~A., \& Kochanek, C.~S.\ 2010, arXiv:1002.3126 
\bibitem[Cenko et al.(2009)]{2009arXiv0911.3150C} Cenko, S.~B., et al.\ 
2009, arXiv:0911.3150 
\bibitem[Churazov et al.(2001)]{2001MNRAS.321..759C} Churazov, E., 
Gilfanov, M., \& Revnivtsev, M.\ 2001, \mnras, 321, 759 
\bibitem[Collier
\& Peterson(2001)]{2001ApJ...555..775C} Collier, S., \& Peterson, B.~M.\ 2001, \apj, 555, 775
\bibitem[Croom et al. (2009)]{2009MNRAS...399..1755} 
Croom, S.M., Richards, G.T., Shanks, T., et al. 2009, MNRAS, 399, 1755
\bibitem[Denney et al.(2009)]{2009ApJ...692..246D} Denney, K.~D., Peterson, 
B.~M., Dietrich, M., Vestergaard, M., \& Bentz, M.~C.\ 2009, \apj, 692, 246 
\bibitem[de Vries et al.(2005)]{2005AJ....129..615D} de Vries, W.~H.,
Becker, R.~H., White, R.~L., \& Loomis, C.\ 2005, \aj, 129, 615
\bibitem[Emmanoulopoulos et al.(2010)]{2010arXiv1001.2045E} 
Emmanoulopoulos, D., McHardy, I.~M., \& Uttley, P.\ 2010, arXiv:1001.2045 
\bibitem[Frank et al.(2002)]{2002apa..book.....F} Frank, J., King, A., 
\& Raine, D.~J.\ 2002, \emph{Accretion Power in Astrophysics} (Cambridge, UK: Cambridge University Press)
\bibitem[Fukugita et al.(1996)]{F96}Fukugita, M., Ichikawa, T., Gunn, J.E., Doi, M., Shimasaku, K., \&
Schneider, D.P. 1996, AJ, 111, 1748
\bibitem[Geha et al.(2003)]{geh03} Geha, M., et al.\ 2003, \aj, 125, 1
\bibitem[Gotz et al.(2009)]{2009GCN..9649....1G} Gotz, D., Mereghetti, S., 
von Kienlin, A., \& Beck, M.\ 2009, GRB Coordinates Network, 9649, 1 
\bibitem[Gunn {\em et al.} 1998]{Gunnetal}Gunn, J.E., et al.\ 1998, AJ, 116, 3040
\bibitem[Giveon et al.(1999)]{giv99} Giveon, U., Maoz, D.,
Kaspi, S., Netzer, H., \& Smith, P.~S.\ 1999, \mnras, 306, 637
\bibitem[Hawkins(1993)]{haw93} Hawkins, M.~R.~S.\ 1993, \nat,
366, 242
\bibitem[Hawkins(2007)]{haw07} Hawkins, M.~R.~S.\ 2007, \aap, 462, 581
\bibitem[Honscheid et al.(2008)]{2008arXiv0810.3600H} Honscheid, K., DePoy, 
D.~L., \& for the DES Collaboration 2008, arXiv:0810.3600
\bibitem[Hook et al.(1994)]{1994MNRAS.268..305H} Hook, I.~M., McMahon,
R.~G., Boyle, B.~J., \& Irwin, M.~J.\ 1994, \mnras, 268, 305
\bibitem[]{} Horne, J.H., \& Baliunas, S.L. 1986, ApJ, 302, 757 
\bibitem[Hughes et al.(1992)]{hug92} Hughes, P.~A., Aller,
H.~D., \& Aller, M.~F.\ 1992, \apj, 396, 469
\bibitem[Ivezi\'{c} et al.(2002)]{2002AJ....124.2364I} Ivezi\'{c}, {\v Z}., 
et al.\ 2002, \aj, 124, 2364 
\bibitem[Ivezi\'{c} et al.(2004)]{ive04} Ivezi\'c, \v Z., et al.\
2004, The Interplay Among Black Holes, Stars and ISM in Galactic Nuclei,
222, 525 [I04]
\bibitem[Ivezi\'{c} et al.(2007)]{2007AJ....134..973I} Ivezi\'{c}, {\v Z}., 
et al.\ 2007, \aj, 134, 973 
\bibitem[Ivezi\'{c} et al.(2008)]{2008arXiv0805.2366I} Ivezi\'{c}, {\v Z}., Tyson, 
J.~A., Allsman, R., Andrew, J., Angel, R., 
\& for the LSST Collaboration 2008, arXiv:0805.2366 
\bibitem[Kaiser et al.(2002)]{2002SPIE.4836..154K} Kaiser, N., et al.\ 
2002, \procspie, 4836, 154
\bibitem[Kawaguchi et al.(1998)]{kaw98} Kawaguchi, T.,
Mineshige, S., Umemura, M., \& Turner, E.~L.\ 1998, \apj, 504, 671
\bibitem[Kelly(2007)]{2007ApJ...665.1489K} Kelly, B.~C.\ 2007, \apj, 665, 
1489 
\bibitem[Kelly et al.(2009)]{2009ApJ...698..895K} Kelly, B.~C., Bechtold,
J., \& Siemiginowska, A.\ 2009, \apj, 698, 895 [KBS09]
\bibitem[Kimball 
\& Ivezi\'{c}(2008)]{2008AJ....136..684K} Kimball, A.~E., \& Ivezi\'{c}, {\v Z}.\ 2008, \aj, 136, 684 
\bibitem[Kollmeier et al.(2006)]{2006ApJ...648..128K} Kollmeier, J.~A., et 
al.\ 2006, \apj, 648, 128 
\bibitem[Koz{\l}owski \& Kochanek(2009)]{2009ApJ...701..508K}
  Koz{\l}owski, S., \& Kochanek, C.~S.\ 2009, \apj, 701, 508
\bibitem[Koz{\l}owski et al.(2010)]{Koz10} Koz{\l}owski, S.,
et al.\ 2010a, \apj, 708, 927 [Koz10]
\bibitem[Koz{\l}owski et al.(2010)]{2010arXiv1002.3365K} Koz{\l}owski, S.,
  et al.\ 2010b, arXiv:1002.3365 
\bibitem[Li
\& Cao(2008)]{2008MNRAS.387L..41L} Li, S.-L., \& Cao, X.\ 2008, \mnras, 387, L41
\bibitem[Liu et al.(2008)]{2008ApJ...677..884L} Liu, H.~T., Bai, J.~M., 
Zhao, X.~H., \& Ma, L.\ 2008, \apj, 677, 884 
\bibitem[]{} Lomb, N.R. 1976, Ap\&SS, 39, 447
\bibitem[MacLeod et al.(2008)]{mac08} MacLeod, C.,
Ivezi\'{c}, {\v Z}., de Vries, W., Sesar, B.,
\& Becker, A.\ 2008, American Institute of Physics Conference Series, 1082, 282
\bibitem[Marconi et al.(2008)]{2008ApJ...678..693M} Marconi, A., Axon, 
D.~J., Maiolino, R., Nagao, T., Pastorini, G., Pietrini, P., Robinson, A., 
\& Torricelli, G.\ 2008, \apj, 678, 693 
\bibitem[Markwardt et al.(2009)]{2009GCN..9645....1M} Markwardt, C.~B., 
Gavriil, F.~P., Palmer, D.~M., Baumgartner, W.~H., 
\& Barthelmy, S.~D.\ 2009, GRB Coordinates Network, 9645, 1 
\bibitem[Martini 
\& Schneider(2003)]{2003ApJ...597L.109M} Martini, P., \& Schneider,
  D.~P.\ 2003, \apjl, 597, L109 
\bibitem[Matthews
\& Sandage(1963)]{1963ApJ...138...30M} Matthews, T.~A., \& Sandage, A.~R.\ 1963, \apj, 138, 30
\bibitem[McHardy(2010)]{2010LNP...794..203M} McHardy, I.\ 2010, Lecture 
Notes in Physics, Berlin Springer Verlag, 794, 203 
\bibitem[Pereyra et al.(2006)]{2006ApJ...642...87P} Pereyra, N.~A., Vanden 
Berk, D.~E., Turnshek, D.~A., Hillier, D.~J., Wilhite, B.~C., Kron, R.~G., 
Schneider, D.~P., \& Brinkmann, J.\ 2006, \apj, 642, 87 
\bibitem[Peterson et al.(2005)]{2005ApJ...632..799P} Peterson, B.~M., et
al.\ 2005, \apj, 632, 799
\bibitem[Press et al.(1992)]{1992ApJ...385..404P} Press, W.~H.,
  Rybicki, G.~B., \& Hewitt, J.~N.\ 1992, \apj, 385, 404
\bibitem[Rengstorf et al.(2006)]{2006AJ....131.1923R} Rengstorf, A.~W.,
Brunner, R.~J., \& Wilhite, B.~C.\ 2006, \aj, 131, 1923
\bibitem[Rybicki \& Press(1992)]{1992ApJ...398..169R} Rybicki, G.~B., \& Press, W.~H.\ 1992, \apj, 398, 169
\bibitem[Rybicki \& Press(1994)]{1994comp.gas..5004R} Rybicki, G.~B., \& Press, W.~H.\ 1994, Computer, 5004
\bibitem[]{} Scargle, J.D. 1982, ApJ, 263, 835
\bibitem[Schmidt et al.(2010)]{2010arXiv1002.2642S} Schmidt, K.~B., 
Marshall, P.~J., Rix, H.-W., Jester, S., Hennawi, J.~F., 
\& Dobler, G.\ 2010, arXiv:1002.2642
\bibitem[Schneider et al.(2007)]{Sch2007} Schneider, D.~P.,  et~al.\ 2007, AJ, 134, 102
\bibitem[Schweitzer et al.(2006)]{2006ApJ...649...79S} Schweitzer, M., et 
al.\ 2006, \apj, 649, 79 
\bibitem[Sesar et al.(2006)]{ses06} Sesar, B., et al.\ 2006,
AJ, 131, 2801
\bibitem[Sesar et al.(2007)]{2007AJ....134.2236S} Sesar, B., et al.\ 2007,
\aj, 134, 2236
\bibitem[Shen et al.(2008)]{2008ApJ...680..169S} Shen, Y., Greene, J.~E.,
Strauss, M.~A., Richards, G.~T., \& Schneider, D.~P.\ 2008, \apj, 680, 169
\bibitem[Smith et al.(2002)]{2002AJ....123.2121S} Smith, J.~A., et al.\
2002, \aj, 123, 2121
\bibitem[Timmer 
\& Koenig(1995)]{1995A&A...300..707T} Timmer, J., \& Koenig, M.\ 1995, \aap, 300, 707 
\bibitem[Tr{\`e}vese et al.(2001)]{tre01} Tr{\`e}vese, D.,
Kron, R.~G., \& Bunone, A.\ 2001, \apj, 551, 103
\bibitem[Tr{\`e}vese
\& Vagnetti(2002)]{tre02} Tr{\`e}vese, D., \& Vagnetti, F.\ 2002, \apj, 564, 624
\bibitem[Udalski et al.(1997)]{Uda97} Udalski et al.\ 1997, Acta Astron 47 319
\bibitem[Udalski et al.(2008)]{Uda08} Udalski et al.\ 2008, Acta Astron 58 69
\bibitem[Uttley et al.(2002)]{2002MNRAS.332..231U} Uttley, P., McHardy, 
I.~M., \& Papadakis, I.~E.\ 2002, \mnras, 332, 231 
\bibitem[Vanden Berk et al.(2004)]{2004ApJ...601..692V} Vanden Berk, D.~E.,
et al.\ 2004, \apj, 601, 692 [VB04]
\bibitem[Vestergaard 
\& Peterson(2006)]{2006ApJ...641..689V} Vestergaard, M., \& Peterson, B.~M.\ 2006, \apj, 641, 689
\bibitem[Voges et 
al.(1999)]{1999A&A...349..389V} Voges, W., et al.\ 1999, \aap, 349, 389 
\bibitem[White et al.(1997)]{1997ApJ...475..479W} White, R.~L., Becker, 
R.~H., Helfand, D.~J., \& Gregg, M.~D.\ 1997, \apj, 475, 479 
\bibitem[Wilhite et al.(2005)]{2005ApJ...633..638W} Wilhite, B.~C., Vanden
Berk, D.~E., Kron, R.~G., Schneider, D.~P., Pereyra, N., Brunner, R.~J.,
Richards, G.~T., \& Brinkmann, J.~V.\ 2005, \apj, 633, 638
\bibitem[Wilhite et al.(2006)]{2006ApJ...641...78W} Wilhite, B.~C., Vanden
Berk, D.~E., Brunner, R.~J., \& Brinkmann, J.~V.\ 2006, \apj, 641, 78
\bibitem[Wilhite et al.(2008)]{2008MNRAS.383.1232W} Wilhite, B.~C.,
Brunner, R.~J., Grier, C.~J., Schneider, D.~P.,
\& Vanden Berk, D.~E.\ 2008, \mnras, 383, 1232
\bibitem[Wold et al.(2007)]{2007MNRAS.375..989W} Wold, M., Brotherton,
M.~S., \& Shang, Z.\ 2007, \mnras, 375, 989
\bibitem[York  et al.(2000)]{York} York, D.G. et al.\ 2000, AJ, 120,
  1579
\end{thebibliography}
\end{document}